\documentclass[times,10pt,conference]{IEEEtran}
\pdfoutput=1
\usepackage[pdftex]{graphicx}
\usepackage{subcaption}
\usepackage[english]{babel}
\usepackage{fancybox}
\usepackage{fancyref}
\usepackage{epic}
\usepackage{times}
\usepackage{epsfig}
\usepackage[hyphens]{url}
\usepackage{multirow}
\usepackage{comment}
\usepackage{setspace}
\usepackage{xspace}
\usepackage{listings}
\usepackage{bibspacing}
\usepackage{amsmath}
\usepackage{fancyvrb}
\usepackage{xspace} % allows new commands to have an extra space after them
\usepackage{hyperref}
\usepackage{color}
\usepackage{enumitem}
\usepackage{gensymb}
\usepackage{amsfonts}
\usepackage{balance}

\usepackage{booktabs}

\usepackage{bbding}

\definecolor{dkgreen}{rgb}{0,0.6,0}
\definecolor{gray}{rgb}{0.5,0.5,0.5}
\definecolor{mauve}{rgb}{0.58,0,0.82}

\definecolor{dollarbill}{rgb}{0.52,0.73,0.4} 

\definecolor{amber}{rgb}{1.0, 0.75, 0.0}
\definecolor{ambersaeece}{rgb}{1.0, 0.49, 0.0}
\definecolor{amaranth}{rgb}{0.9, 0.17, 0.31}

\hypersetup{
  colorlinks,
  citecolor=blue,
  linkcolor=blue,
  urlcolor=blue}

\renewenvironment{thebibliography}[1]{%
  \begin{oldthebibliography}{#1}%
    \setlength{\parskip}{.35ex}%
    \setlength{\itemsep}{0ex}%
  }%
  {%
  \end{oldthebibliography}%
}

\newcounter{packednmbr}

\def\dsp{\def\baselinestretch{1}\large\normalsize}
\dsp

\def\ffsp{\def\baselinestretch{.85}\large\normalsize}

\date{} 
\title{Are We Susceptible to Rowhammer?\\An End-to-End Methodology for Cloud Providers}
\author{Lucian Cojocar, Jeremie Kim$^{\S\dag}$, Minesh Patel$^\S$,
  Lillian Tsai$^\ddag$,\\
  Stefan Saroiu, Alec Wolman, and Onur Mutlu$^{\S\dag}$\\
  Microsoft Research, $^\S$ETH Z\"urich, $^\dag$CMU, $^\ddag$MIT}

\begin{document}
\maketitle

\setstretch{0.93}

\noindent
\begin{abstract}
  Cloud providers are concerned that Rowhammer poses a potentially critical
  threat to their servers, yet today they lack a systematic way to test
  whether the DRAM used in their servers is vulnerable to Rowhammer
  attacks.  This paper presents an end-to-end methodology to determine if
  cloud servers are susceptible to these attacks. With our methodology, a
  cloud provider can construct worst-case testing conditions for DRAM.
  
  We apply our methodology to three classes of servers from a major cloud
  provider.  Our findings show that none of the CPU instruction sequences used in
  prior work to mount Rowhammer attacks create worst-case DRAM testing
  conditions. To address this limitation, we develop an instruction
  sequence that leverages microarchitectural side-effects to ``hammer''
  DRAM at a near-optimal rate on modern Intel Skylake and Cascade Lake platforms.
  We also design a DDR4 fault injector that can reverse engineer row
  adjacency for any DDR4 DIMM. When applied to our cloud provider's DIMMs,
  we find that DRAM rows do not always follow a linear map.
\end{abstract}

\section{Introduction}
\label{sec::intro}

The consequences of a large-scale security compromise of a cloud provider
cannot be overstated. An increasing number of banks, hospitals, stores,
factories, and universities depend upon cloud resources for their
day-to-day activities. Users store important and private data in the cloud,
including tax returns, health records, e-mail, and backups. Home devices
and home automation are also becoming reliant on cloud infrastructure. An
attack that steals or deletes data, or performs a large-scale denial of
service (DoS) attack on the cloud, would be catastrophic to both cloud
providers and customers.

Today's DRAM is vulnerable to memory disturbance errors: a high rate of
accesses to the same address in DRAM flips bits in data stored in nearby
addresses~\cite{kim2014disturbance}. Rowhammer attacks generate adversarial
workloads that exploit disturbance errors to flip the value of
security-critical bits, such as an OS's page table
entries~\cite{kaveh2016flip-feng-shui,google2015projectzero}, a browser's
isolation sandbox~\cite{google2015projectzero}, a CPU's isolation
mechanism~\cite{jang2017sgx}, an encryption
key~\cite{kaveh2016flip-feng-shui}, or a CPU instruction
opcode~\cite{gruss2018anotherflip}. Even worse, mounting Rowhammer requires
no special privilege: attacks have been demonstrated by running user-level
code~\cite{google2015projectzero,kaveh2016flip-feng-shui}, JavaScript
programs~\cite{bosman2016dedup-est-machina,gruss2016rowhammer-js}, and even
by sending RDMA network
packets~\cite{tatar2018throwhammer,lipp2018nethammer}.  Because all DRAM is
potentially susceptible, Rowhammer attacks are an easy way to mount a
large-scale catastrophic attack on the cloud.  The combination of
easy-to-mount and easy-to-scale makes Rowhammer a formidable potential
attack vector to the cloud.

Unfortunately, the threat that Rowhammer poses to cloud providers remains
unclear. Security researchers have been publishing a stream of
proof-of-concept exploits using Rowhammer that affect all types of DRAM,
including DDR3~\cite{kim2014disturbance,google2015projectzero},
DDR4~\cite{veen2016drammer,lanteigne2016thirdio}, and ECC-equipped
DRAM~\cite{cojocar2019ecc}. DRAM vendors claim that their memory is safe
against Rowhammer attacks; these claims are delivered to cloud providers
with each new DRAM feature: DDR4~\cite{lanteigne2016thirdio}, ECC-equipped
DRAM~\cite{greenberg2015SafeDDR4,goodin2016DDR4}, and TRR-equipped
DRAM~\cite{lee2014rhfree,greenberg2015SafeDDR4}. There is a large gap
between a proof-of-concept exploit carried out in a research lab and an
actual attack in the wild. In fact, no evidence indicates that Rowhammer
attacks have been carried out in practice.  In the absence of attacks in
the wild, one could easily dismiss Rowhammer as a credible threat.

This confusion is further fed by the lack of a systematic methodology to
test for Rowhammer. Previous proof-of-concept attacks used varied
methodologies to mount Rowhammer~\cite{kim2014disturbance,
  google-rh-blackhat, google2015projectzero, xiao2016cloudflops,
  bosman2016dedup-est-machina, qiao2016new, bhattacharya2016curious,
  bhattacharya2018advanced, gruss2016rowhammer-js, pessl2016drama,
  kaveh2016flip-feng-shui, veen2016drammer, lanteigne2016thirdio,
  jang2017sgx, aga2017good, gruss2017guide, gruss2018anotherflip,
  tatar2018defeating, frigo2018glitch-vu, tatar2018throwhammer,
  lipp2018nethammer, poddebniak2018attacking, carre2018openssl,
  cojocar2019ecc, barenghi2018software, zhang2018triggering,
  fournaris2017exploiting} based on heuristics, without
rigorously characterizing their effectiveness.  While such approaches
can demonstrate an attack's viability, they are unsuitable for testing
purposes because they cannot distinguish between Rowhammer-safe memory and
a sub-optimal, imperfect testing methodology. Lacking a comprehensive
testing tool, cloud providers find it difficult to ascertain the degree to
which Rowhammer poses a threat to their infrastructure.

Building a systematic and scalable testing methodology must overcome two
serious practical challenges. First, it must devise a sequence of CPU
instructions that leads to a maximal rate of row activations in DRAM.  This
sequence must overcome the hardware's attempts to capture locality, such as
the CPU's re-ordering of instructions and the DRAM controller's re-ordering
of memory accesses. For this, we need to measure the row activation rates
of the instruction sequences used by previous work, identify their
bottlenecks, and test new candidates that overcome these bottlenecks.
Previous work showed that the probability of flipping bits in a Rowhammer
attack increases with the rate of row
activations~\cite{mutlu2019rowhammer,kim2014disturbance}.

The second challenge is determining row adjacency in a DRAM device.
Contiguous virtual addresses do not map linearly to DRAM rows and are in
fact subject to three mapping layers. The OS maintains a
virtual-to-physical address map that can change often and at
runtime~\cite{irazoqui2015cacheSlice,islam2019spoiler}; if present,
virtualization adds another mapping layer due to guest-physical addresses.
The memory controller further maps physical addresses to logical bus
addresses specified in terms of ranks, banks, rows, and
columns~\cite{kim2016ramulator, ramulatorgithub}.  The final mapping is
done by the DRAM device itself, where a device can remap adjacent logical
bus addresses to non-adjacent physical rows~\cite{kim2012case}.  DRAM
vendors consider these maps to be trade secrets and strongly guard their
secrecy.

Prior techniques to reverse engineer row adjacency with a commodity memory
controller \emph{rely} on Rowhammer attacks~\cite{schwarz2016drama,
  pessl2016drama, tatar2018defeating} and work only on DIMMs that succumb
to them. Once bits flip, the flips' locations
reveal information on row adjacency.  This creates a chicken-and-egg
problem: testing DIMMs' resiliency to Rowhammer requires knowing row
adjacency information, and reverse engineering row adjacency requires
having DIMMs succumb to Rowhammer.

This paper presents solutions to both challenges and combines them in an
end-to-end methodology that creates worst-case conditions for testing the
presence of disturbance errors in DRAM. Our methodology lets cloud
providers construct (1) an instruction sequence that maximizes the rate of
DRAM row activations on a given system, and (2) accurate maps of address
translations used by the system's hardware. Armed with this knowledge, a
cloud provider can develop a quality control pipeline to test its servers'
DRAM and ultimately characterize the risk of a Rowhammer attack to its
infrastructure.

We start by showing how a memory bus analyzer can characterize the
\emph{effectiveness} of a sequence of CPU instructions when hammering
memory. It can measure the rate of activation commands and compare them to
the \emph{optimal} rate (i.e., the highest rate of activations to memory
that the specifications allow). Our results show that all instruction
sequences used in previous work hammer memory at a sub-optimal rate. Most
previous sequences have a rate that is at most half of optimal, and the
most effective previous sequence is 33\% from optimal. We tested 42
different instruction sequences, including those found in previous
work~\cite{kim2014disturbance, google2015projectzero, xiao2016cloudflops,
  bosman2016dedup-est-machina, qiao2016new, bhattacharya2016curious,
  gruss2016rowhammer-js, pessl2016drama, kaveh2016flip-feng-shui,
  veen2016drammer, lanteigne2016thirdio, jang2017sgx, aga2017good,
  gruss2017guide, gruss2018anotherflip, tatar2018defeating}, and developed
additional variants.

Our characterization sheds light on the factors that prevent these
instruction sequences from having a high rate of activations. One
significant factor is out-of-order execution -- the CPU constantly
re-orders memory accesses to increase the likelihood they are served from
the cache.  \emph{Out-of-order execution can act as a de facto rate limiter
  to Rowhammer attacks.}  Equally significant are memory barriers. Some
instruction sequences use memory barriers to order their memory accesses.
Although barriers do prevent out-of-order execution, we find they are too
slow.  \emph{Instruction sequences that use memory barriers lack the
  performance necessary to create a high rate of activations.}

This analysis led us to construct a near-optimal instruction sequence that
maximizes the rate of activations, effectively matching the minimum row
cycle time of the DDR4 JEDEC~\cite{ddr4jedec} spec. Our instruction
sequence differs considerably from all sequences used in previous work
because it uses no explicit memory accesses (e.g., no load or store
instructions).  Instead, we craft our instruction sequence to leverage
microarchitectural side-effects of \emph{clflushopt} that issues memory
loads in order and without the use of memory barriers.

We overcome our second challenge, determining row adjacency, by designing
and building a DDR4 fault-injector that \emph{guarantees} that any DIMM
succumbs to a Rowhammer attack. Our fault-injector is both low-cost and
compatible with any DDR4 motherboard.  It suppresses all refresh commands
received by a DIMM for a fixed period of time. The absence of refreshes
ensures the success of a Rowhammer attack in flipping bits on today's DDR4
memory. The location and density of bit flips lets our methodology reverse
engineer the physical row adjacency of \emph{any} DDR4 DRAM device. To our
knowledge, \emph{ours is the first fault injector capable of injecting
  faults into DDR4 commands.}

We leverage the fault injector to reverse engineer physical adjacency in a
major cloud provider's DRAM devices supplied by three different vendors.
Our results show that logical rows do not always map linearly, but instead
can follow a \emph{half-row} pattern, where two halves of a single row are
adjacent to different rows. A methodology that uses guess-based heuristics
to determine row adjacency will be ineffective in testing these half-row
patterns. We also find that mounting a Rowhammer attack on a victim row
that follows a half-row pattern requires hammering more aggressor rows than
it does for a victim row that is contiguous within a single physical row.

We applied our methodology on a major cloud provider's three most recent
classes of servers based on Intel's Cascade Lake, Skylake, and Broadwell
architectures. On the two newest architectures, Cascade Lake and Skylake,
our methodology achieves a near-optimal rate of activations by using
\emph{clflushopt} to ``hammer'' memory, an instruction introduced with the
Skylake architecture.  Finally, we used the fault injector to successfully
reverse engineer physical row adjacency on all three classes of servers.

%\input{landscape}
%\section*{[Section Removed] The Confusing Landscape of the Rowhammer Threat}

\section{Background}
\label{sec::background}

Rowhammer bit flips result from circuit-level charge leakage mechanisms
that are exacerbated by certain memory access patterns. This section
provides a high-level background on DRAM and the physical mechanisms
responsible for the Rowhammer phenomenon in order to facilitate the
understanding of our work. More detail on Rowhammer and its system-level
implications can be found in~\cite{kim2014disturbance, onur-date17,
  mutlu2019rowhammer}.

\subsection{DRAM Organization}

\begin{figure}[t]
\begin{center}
\includegraphics[width=0.48\textwidth]{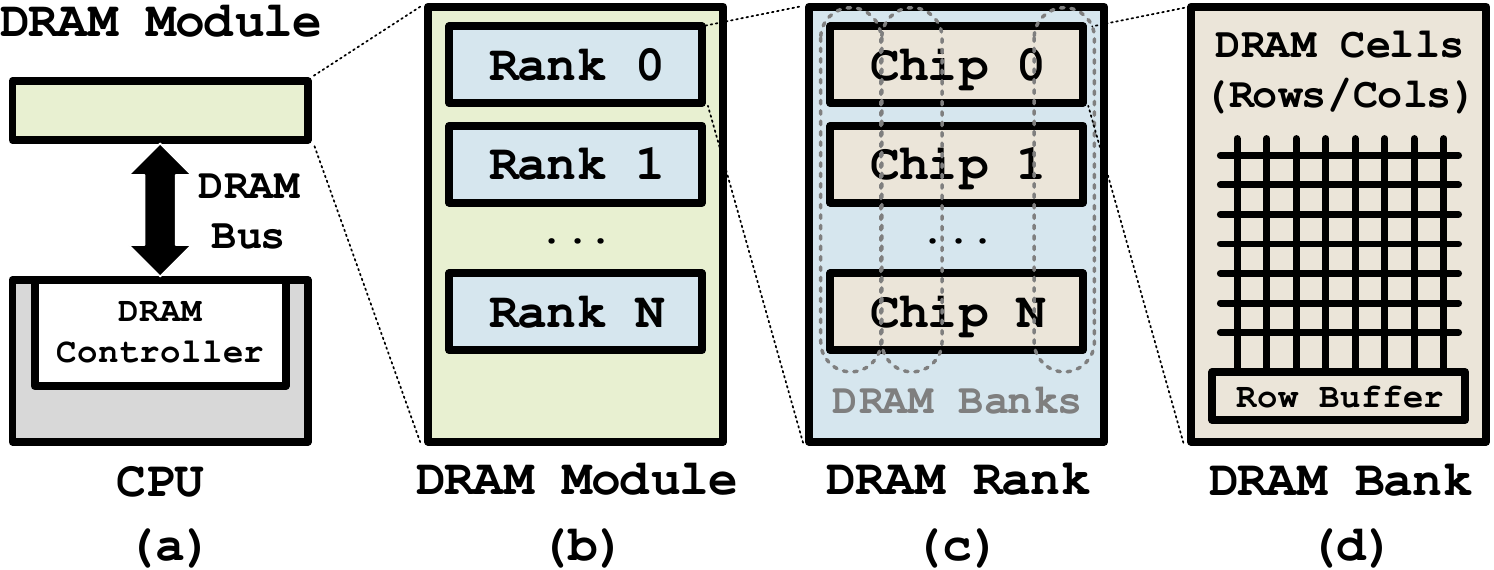}
\end{center}
\vspace{-.3cm}
\caption{\ffsp Typical DRAM organization.}
\vspace{-.5cm}
\label{fig:dram_org}
\end{figure}

DRAM comprises a hierarchy of two-dimensional arrays, as shown in
Figure~\ref{fig:dram_org}. At the top level, a \emph{DRAM controller}
interfaces with a \emph{DRAM rank} over a \emph{channel}
(Figure~\ref{fig:dram_org}a). The channel conveys \emph{DRAM commands},
\emph{addresses}, and \emph{data} between the DRAM controller and the DRAM
rank. In modern systems, multiple DRAM ranks are typically combined in a
\emph{DRAM module} (Figure~\ref{fig:dram_org}b). The DRAM
controller uses \emph{chip-select} signals to interface with only a single
DRAM rank at any given time.

A DRAM rank consists of multiple physical \emph{DRAM chips}
(Figure~\ref{fig:dram_org}c).  The DRAM controller is unaware of
how a single rank is partitioned into individual DRAM chips.  Instead, it
sees each rank as the union of multiple banks that are each striped
across the physical DRAM chips that form the rank. Thus, one
bank spans \emph{multiple} DRAM chips, and a single DRAM chip stores data
from multiple banks. This has implications on how failures affect banks.
Different DRAM chips can have different failure profiles depending on how
they are manufactured; thus, a ``weak'' DRAM chip affects \emph{multiple}
banks. However, only a portion of each bank is affected, namely, the
portion that corresponds to the weak DRAM chip.

% A DRAM rank consists of multiple physical \emph{DRAM chips}, each of which
% is subdivided into multiple physical \emph{DRAM banks}
% (Figure~\ref{fig:dram_org}c).  However, the DRAM controller is unaware of
% how a single rank is partitioned into individual DRAM chips. Instead, the
% DRAM controller sees each rank as the union of multiple \emph{logical}
% banks, each of which is formed by striping data across the corresponding
% \emph{physical} bank of each chip in the rank.

% The data of a single DRAM bus transfer is
% transparently striped across DRAM chips.

% , and the DRAM controller observes each bank across
% all of the physical chips as one logical bank.

DRAM banks within a chip are further subdivided into \emph{rows} and
\emph{columns} of storage cells (Figure~\ref{fig:dram_org}d), where each
cell encodes a single bit of data using the amount of charge stored in a
capacitor (i.e., data ``1'' as either fully-charged or fully-discharged,
and data ``0'' as its opposite). The DRAM controller accesses a cell by
specifying a \emph{row address} and a \emph{column address} to a particular
bank. It has no knowledge of the physical layout of banks or that a bank
comprises multiple physical chips.

\subsection{How DRAM Operates}

A DRAM read operation \emph{senses} the amount of charge stored in cell
capacitors. It is subdivided into three phases: \textbf{(1) Activation
  (ACT):} the charge stored in an entire DRAM row's cells within a bank is
sensed and stored in a \emph{row buffer}. The row buffer acts as a fast
cache; subsequent accesses to the same row do not require another
activation. Instead, data is read out of the row buffer. \textbf{(2) Read
  (RD):} the contents of the row buffer at a given column address are
returned to the DRAM controller. \textbf{(3) Precharge (PRE):} the bank is
prepared for the next access by disconnecting and clearing the active row
buffer.

All rows in a bank share one row buffer. Within a bank, the DRAM controller
activates and reads from only one row at a time. Write operations work
similarly to reads.

\textbf{Refresh (REF):} DRAM cells leak their charge, which can cause data
loss when cells are not accessed frequently enough. To prevent this, the
DRAM controller issues periodic refresh operations that replenish the
cells' charge.  The DDR4 standard specifies that 8192 refresh commands
be issued during a 64 ms time period~\cite{ddr4jedec},
which results in considerable power consumption and performance
overhead~\cite{liu2012raidr}.

\subsection{The Rowhammer Threat}

Modern deeply-scaled DRAM devices are susceptible to \emph{Rowhammer} -- a
circuit-level disturbance phenomenon in which repeatedly activating and
precharging a given row accelerates charge leakage in the storage cells of
physically nearby rows~\cite{kim2014disturbance}. With enough
activate/precharge cycles, this phenomenon causes bit flips.

Prior work extensively studied the statistical characteristics of
Rowhammer-susceptible storage cells~\cite{kim2014disturbance,
  park2016statistical, lanteigne2016thirdio, park2016root} and found that
the magnitude of the error rate depends significantly on the number of
activate/precharge cycles.  Other factors affecting the error rates include
the particular device under test, the ambient temperature, and the data
pattern written into the cells.  Recent work~\cite{yang2019trap} identified
the precise charge leakage path responsible for Rowhammer errors and
provided a detailed model that explains experimental observations made by
previous work~\cite{park2016statistical}.

Proposals for mitigating and/or preventing Rowhammer errors abound in both
academia~\cite{anvil, moin-rowhammer, seyedzadeh2017counter,
  brasser2017can, irazoqui2016mascat, son2017making, gomez2016dram,
  van2018guardion, lee2018twice, bu2018srasa, bains2015row, bains14d,
  bains14c, bains16refresh-cmd, greenfield15condition-monitoring,
  konoth2018zebram, izzo2017reliably, gong2018memory, jones2017holistic,
  kline2017sustainable, schilling2018pointing, vig2018rapid} and
industry~\cite{rh-apple, rh-hp, rh-lenovo, rh-cisco};
see~\cite{mutlu2019rowhammer} for a detailed survey of these works.
However, while DRAM manufacturers claim that modern DRAM devices are
resilient to Rowhammer
bit-flips~\cite{lee2014rhfree,greenberg2015SafeDDR4}, it is unclear what
causes this resilience and under what conditions the Rowhammer-prevention
mechanism may fail. Furthermore, even if a particular DRAM device is
supposedly protected, there is no known way to verify that it is in fact
fully resistant to Rowhammer bit flips.

\subsection{Intel-based Cloud Server Architectures}

For the cloud market, Intel offers the {\em scalable-performance} variant
of Xeon processors (often referred to as Xeon-SP). SP indicates a
server class CPU for multi-socket cloud motherboards, introduced with
Intel Skylake. Intel Broadwell uses EP to designate multi-socket server CPUs.

We performed our experiments on the three most recent generations of Xeon
servers: Broadwell-EP, Skylake-SP, and Cascade Lake-SP.  Skylake is a major
architectural revision of Broadwell. Some of our results are affected by
these architectural differences, such as whether the architecture supports
the \emph{clflushopt} instruction.  Cascade Lake is a minor revision of
Skylake, and, indeed, our results are similar on both of these platforms.
Intel announced the upcoming release of Ice Lake-SP, a major architectural
revision of Skylake, but these CPUs are not available at this time.

\section{Challenges of DRAM Testing}
\label{sec::challenges}

Previous work established a direct relationship between the number of DRAM
row activations within back-to-back refresh (REF) commands and the number of observed
bit flips. A good example is Figure 2 in a recent
paper~\cite{mutlu2019rowhammer}. This observation is not new; it
goes back to the original paper showing DRAM disturbance
errors~\cite{kim2014disturbance}. To test a DRAM row for its susceptibility
to Rowhammer attacks, we repeatedly activate two adjacent rows co-located in
the same bank.  Alternating accesses to two rows ensures that each access
first precharges the open row and then activates the row being accessed.
We refer to the rows we activate as \emph{aggressor} rows,
and the tested row as the \emph{victim} row.

Naively, to mount Rowhammer, one would like to activate only a single row
repeatedly. Unfortunately, there is no way to accomplish this in practice
on systems using an \emph{open-page policy}~\cite{kaseridis2011openpage}
(the terms page and row are equivalent in this context).  According to this
policy, the memory controller leaves a DRAM row open in the row buffer
after access. Accessing the same row leads to reading repeatedly from the
bank's row buffer rather than to multiple row activations. Open-page policy
is the default configuration in most systems today, including the servers
used by our cloud provider.

\subsection{Fundamental Testing Requirements}

To identify all possible Rowhammer failures when a system is operational,
our testing methodology must replicate the worst-case Rowhammer testing
conditions. We identify two fundamental testing requirements: \textbf{(1)}
The methodology must activate DRAM rows at the highest possible rate.
Repeatedly activating a row toggles the wordline's voltage, which causes
disturbance errors. The testing methodology must toggle the wordline
voltage at the fastest (i.e., worst-case) rate possible to ensure the
largest number of wordline activations within a refresh interval.
\textbf{(2)} The methodology must test each row by identifying and toggling
\emph{physically adjacent} rows within DRAM. Rowhammer attacks are most
effective when aggressor rows are physically adjacent to the victim
row~\cite{kim2014disturbance}. Hammering rows without precise knowledge of
physical adjacency is \emph{not} an effective testing methodology.

\subsection{Challenges of Generating the Highest Rate of ACT Commands}

The initial study of DRAM disturbance errors~\cite{kim2014disturbance}
directly attached DRAM modules to an FPGA board that acts as the memory
controller and can issue arbitrary DRAM commands~\cite{softmc,softmc-safarigithub}. The FPGA was programmed
to issue ACT commands at the optimal rate determined by the DRAM timing
parameter tRC (i.e., minimum row cycle time) in the JEDEC specification
sheet~\cite{ddr3jedec, ddr4jedec}.

In contrast, testing DRAM on a cloud server is challenging due to the
complexity of modern machines. Instruction sequences execute out-of-order,
their memory accesses interact with a complex cache hierarchy designed to
capture and exploit locality, and memory controllers implement complex DRAM access
scheduling algorithms~\cite{mutlu2008parallelism,usui2016dash}. To comprehensively test a cloud server for
Rowhammer-vulnerable DRAM devices, we need to find the \emph{optimal
  instruction sequence} that, when executed, causes
the memory controller to issue ACT commands at the
optimal rate (every tRC time interval).

%on a cloud server, a CPU's memory controller issues the ACT
%commands. Memory controllers follow the JEDEC specification that includes a
%long list of timing variables controlling how to access the DRAM devices.  The
%DRAM vendors design their DRAM devices to operate correctly when they are
%accessed according to the JEDEC spec. An important parameter is the minimum row
%cycle time (tRC) present in the JEDEC spec~\cite{ddr3jedec,ddr4jedec} because
%it controls the minimum amount of time between two ACT commands. \jk{If we want
%to maximize the effectiveness of a Rowhammer attack, the most effective}
%testing methodology must find the \emph{optimal instruction sequence} that,
%when executed, makes the memory controller issue ACT commands separated by a
%time interval equal to the minimum row cycle time.

Previous work on Rowhammer used a variety of different instruction
sequences to mount the attack~\cite{kim2014disturbance, google-rh-blackhat,
  google2015projectzero, xiao2016cloudflops, bosman2016dedup-est-machina,
  qiao2016new, bhattacharya2016curious, bhattacharya2018advanced,
  gruss2016rowhammer-js, pessl2016drama, kaveh2016flip-feng-shui,
  veen2016drammer, lanteigne2016thirdio, jang2017sgx, aga2017good,
  gruss2017guide, gruss2018anotherflip, tatar2018defeating,
  frigo2018glitch-vu, tatar2018throwhammer, lipp2018nethammer,
  poddebniak2018attacking, carre2018openssl, cojocar2019ecc,
  barenghi2018software, zhang2018triggering, fournaris2017exploiting}. It is unclear
whether these sequences lead to different rates of ACT commands, which
sequence is the most effective, and how far from the optimal ACT rate each sequence is.
Most previous work evaluated the effectiveness of an instruction
sequence mounting a Rowhammer attack via the \emph{number of flipped bits} metric.
Unfortunately, this metric is inadequate for testing DRAM because it fails
to distinguish a case where memory is safe from the case where the instruction
sequence is ineffective.

\subsection{Challenges of Determining Adjacency of Rows Inside a DRAM Device}
\label{sec::methodology::challenges::adjacency}

\begin{figure}[t]
\begin{center}
\includegraphics[height=40mm]{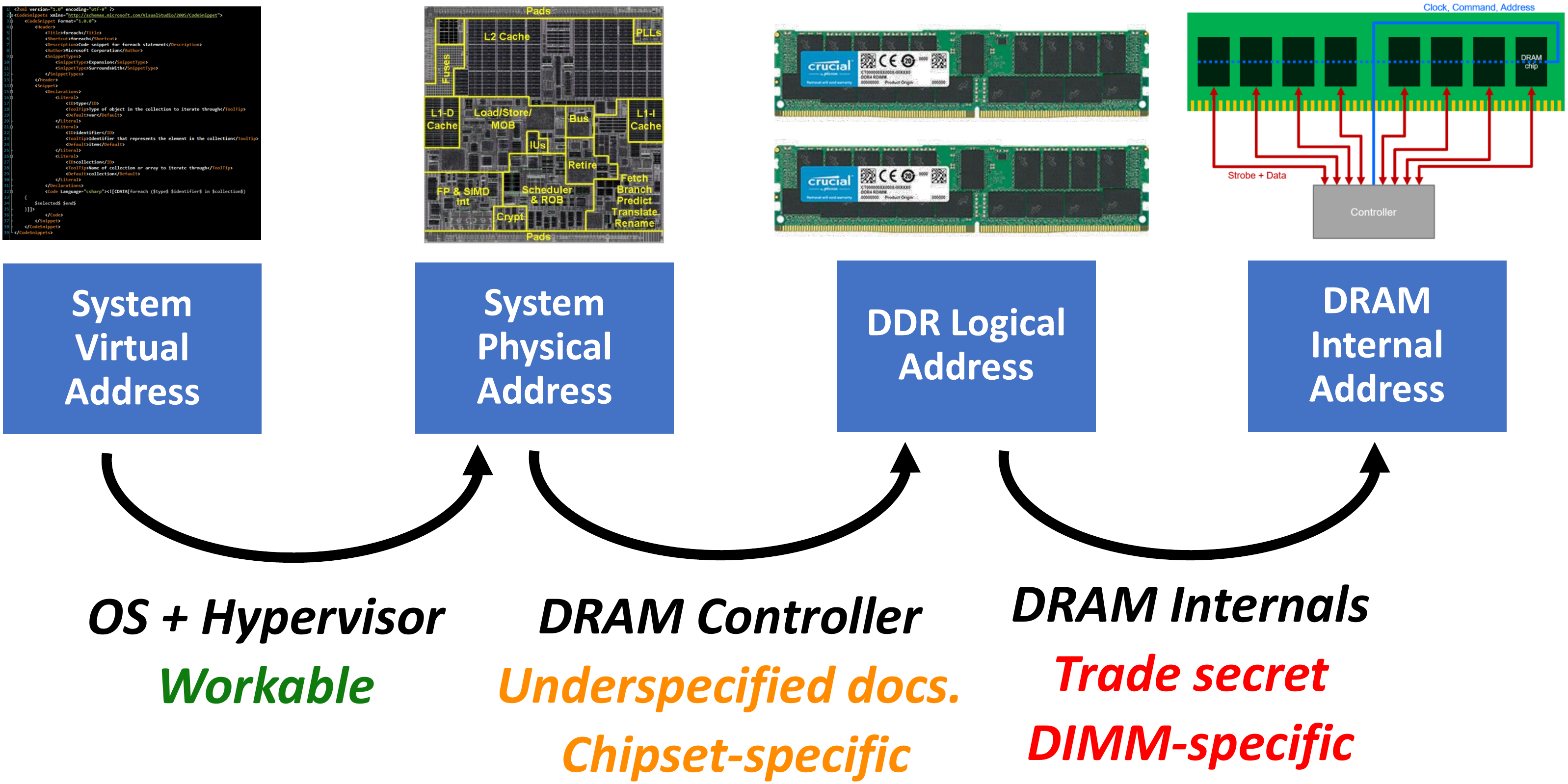}
\end{center}
\vspace*{-.3cm}
\caption{\ffsp Three remapping layers from virtual address to DRAM internal address.}
\label{fig::maps}
\vspace*{-.5cm}
\end{figure}

Instruction sequences access memory via virtual addresses. Virtual
addresses are subject to at least three different remappings until mapped
to an internal set of cells inside a DRAM device.  Figure~\ref{fig::maps}
shows the three different remapping layers.

\noindent
\textbf{1. Virtual-to-Physical:} An OS maintains the map of virtual to
  physical addresses. A virtual address gets translated into a physical
  address by the CPU's Memory Management Unit (MMU). Virtualized cloud
  servers have an additional mapping layer due to virtualization -- virtual
  addresses are first remapped to guest-physical addresses, which are then
  remapped to host physical addresses.

\noindent
\textbf{2. Physical-to-Logical:} A memory controller maintains a map (or
mapping function) of physical addresses to DDR logical addresses (also
called linear memory addresses~\cite{tatar2018defeating}) and translates
incoming physical addresses to DDR logical addresses. These DDR addresses
are specified in terms of channel, DIMM, rank, bank, row, and column.
These maps, seldom public, differ from one CPU architecture to another, and
they are subject to various BIOS settings, such as interleaved
memory~\cite{kim2010atlas}. On Skylake~\cite{intel-skylake} and
Broadwell~\cite{intel-broadwell}, different memory controller
configurations (e.g., page policies)~\cite{gill2010everything} change these
maps.

\noindent
\textbf{3. Logical-to-Internal:} Vendors remap logical addresses in order
to decrease the complexity of internal circuitry and PCB traces because
some maps are electrically easier to build than
others~\cite{liu2013experimental, lee2017diva, khan2016parbor}. Remapping
also lets vendors take advantage of the fact that DRAM devices have
redundancy to tolerate a small number of faults per chip; vendors test for
these faulty elements post packaging and remap wordlines or bitlines to
redundant ones elsewhere in the array (i.e., \emph{post package
  repair})~\cite{futureplus2017ppr, liu2013experimental}.  Memory vendors
regard these maps as trade secrets.

% Prior work used side-channel attacks, reduced timing parameters, thermal
% heaters, physical probing, and Rowhammer attacks to reverse engineer parts
% of these maps~\cite{schwarz2016drama, pessl2016drama,lee2017diva,
%   jung2016crosshair, tatar2018defeating}.  Unfortunately, such techniques
% have shortcomings that prevent our methodology from using them: they are
% coarse-grained~\cite{schwarz2016drama,pessl2016drama,lee2017diva,jung2016crosshair},
% invasive~\cite{jung2016crosshair}, do not capture DRAM internal
% addresses~\cite{schwarz2016drama,pessl2016drama}, or are not
% consistent~\cite{schwarz2016drama, pessl2016drama, tatar2018defeating}.
% Appendix~\ref{appendix::re_mappings_limitations} describes these
% limitations.

\begin{table*}[t]
  \vspace{-.5cm}
  \centering
{\small  \begin{tabular}{@{}lccccc@{}}
\toprule
&\textbf{Side-channels}&\textbf{Reduced
  timings}&\textbf{Heaters}&\textbf{Physical probing}&\textbf{Rowhammer attacks}\\
&\cite{schwarz2016drama,
  pessl2016drama}&\cite{lee2017diva,kim2018solar}&\cite{jung2016crosshair}&\cite{schwarz2016drama,pessl2016drama}&\cite{schwarz2016drama, pessl2016drama, tatar2018defeating}\\
\midrule
\textbf{Fine-grained}&\XSolidBold&\XSolidBold&\XSolidBold&\CheckmarkBold&\CheckmarkBold\\
\textbf{Non-invasive}&\CheckmarkBold&\CheckmarkBold&\XSolidBold&\CheckmarkBold&\CheckmarkBold\\
\textbf{Finds Internal DRAM addresses}&\CheckmarkBold&\CheckmarkBold&\CheckmarkBold&\XSolidBold&\CheckmarkBold\\
\textbf{Consistent}&\CheckmarkBold&\CheckmarkBold&\CheckmarkBold&\CheckmarkBold&\XSolidBold\\
\bottomrule
\end{tabular}}
%\vspace{-.25cm}
\caption{\ffsp Limitations of previous work on reverse engineering row adjacency
  inside DRAM.}
\label{tbl::re_mappings_limitations}
\vspace{-.4cm}
\end{table*}

Previous work used a combination of side-channel attacks, reduced timing
parameters, thermal heaters, physical probing, and Rowhammer attacks to
reverse engineer parts of these maps~\cite{schwarz2016drama,
  pessl2016drama,lee2017diva, jung2016crosshair, tatar2018defeating}.
Unfortunately, such techniques have shortcomings that prevent our
methodology from using them. They are either too
coarse-grained~\cite{schwarz2016drama,pessl2016drama,lee2017diva,jung2016crosshair},
invasive (i.e., potentially damaging the chips)~\cite{jung2016crosshair},
inconsistent~\cite{schwarz2016drama, pessl2016drama, tatar2018defeating},
or they do not capture DRAM internal addresses~\cite{schwarz2016drama,pessl2016drama}.

\noindent
\textbf{Side-channel attacks are coarse-grained.} All memory accesses to
a DRAM bank share one row buffer. Prior work measured two addresses' access
times to determine whether they are co-located in the same
bank~\cite{schwarz2016drama, pessl2016drama}.  Sequentially accessing any
two rows within the same bank takes longer than accessing two rows located
in different banks.  However, this method cannot provide finer-grained
adjacency information.  The time spent accessing two rows sequentially in
the same bank \emph{is unrelated} to the rows' locations within the bank.

\noindent
\textbf{Reduced timing parameters is coarse-grained.} Another technique
uses the distance from a row to the bank's row buffer~\cite{lee2017diva, kim2018solar}.
This technique induces errors by accessing memory using shorter-than-normal
DDR timing values. Data stored in a cell closer to the row buffer has a
shorter distance to travel than data stored further away~\cite{lee2013tiered}, and thus, it has
a lower likelihood to fail. This technique provides coarse-grained and
approximate row adjacency information only.  Adjacent rows have a
negligible difference in access times, and detecting such small differences
is challenging.

\noindent
\textbf{Using heaters is invasive and coarse-grained.} Another technique
surrounds a DIMM with resistive heaters, applies a thermal gradient on each
DRAM device, and conducts a retention error
analysis~\cite{jung2016crosshair}. This approach requires high
temperatures, in excess of $99^{\circ}$C. Cloud providers are reluctant to
adopt a testing methodology that heats up their hardware. Also, the thermal
gradient approach is coarse-grained; it can only determine
\emph{neighborhood relations} rather than \emph{row adjacency}.

\noindent
\textbf{Physical probing does not capture DRAM internal addresses.}
Another approach uses an oscilloscope probe to
capture a DDR electrical signal while issuing memory
accesses~\cite{schwarz2016drama,pessl2016drama}. This approach cannot
reverse engineer how DDR logical addresses map to DRAM internal addresses
(Figure~\ref{fig::maps}).  Previous work used this technique to reverse
engineer only bank addresses~\cite{schwarz2016drama,pessl2016drama}.
Reverse engineering row addresses would incur significant additional effort
for two reasons. First, row addresses require 22 individual probes, whereas
bank addresses require only 4 probes. Second, the signals encoding row addresses
change from one DDR4 command to another (Table~\ref{tbl::ddr4_cmds}).  The
reverse engineering effort would need to ensure that the probes
capture only the signals encoding DDR4 row activation, and not other commands.
In contrast, signals encoding bank addresses are shared by DDR4 row
activation, read, write, and precharge commands. Capturing the signals
corresponding to \emph{any one} of these DDR4 commands reveals the bank
address.

\noindent
\textbf{Rowhammer attacks are not consistent because they may not cause
  failures.} Another technique mounts Rowhammer attacks on every row in
DRAM and correlates each row's density of bit flips with
adjacency~\cite{schwarz2016drama, pessl2016drama, tatar2018defeating}.
Generating a high rate of activations is enough to cause many bit flips
on some DIMMs, but not on all. This approach is unsuitable for testing
memory resilient to Rowhammer. This is an instance of a chicken-and-egg
problem: (1) testing DRAM for Rowhammer susceptibility requires knowing
the adjacency of rows inside DRAM devices, and (2) deducing row adjacency
requires flipping bits using Rowhammer attacks.

Table~\ref{tbl::re_mappings_limitations} summarizes the limitations of
previous work on reverse engineering row adjacency inside DRAM.

\section{Step 1: Generating the Highest Rate of ACT Commands on a Server
Architecture}

We first describe the system setup we used to measure the
rate of row activations of an instruction sequence. Measuring ACT rates
lets us (1) find which instruction sequence generates the highest ACT rate
on a particular server platform, and (2) quantify the difference between
this highest ACT rate and the optimal rate determined from DRAM
datasheets~\cite{ddr4jedec}. We then evaluate the performance of
instruction sequences used by prior work to mount Rowhammer. Finally, we
present a new instruction sequence that generates near-optimal row
activation rates on Intel Skylake and Cascade Lake architectures.

\subsection{System Setup for Measuring ACT Rates}

To determine the instruction sequence that generates the highest rate of
ACT commands, we used the FS2800 DDR Detective from FuturePlus Systems with
two DIMM interposers for DDR3 and DDR4~\cite{ddr_detective}.  This system
can monitor, capture, and measure DDR commands issued over the command bus
to the DIMM using a DDR interposer and an FPGA that triggers on specific
DDR commands or memory addresses. Once triggered, the FPGA records all DDR
commands it observes on the bus and stores them in buffers, which are later
transferred to a host computer over USB.

%XXX: Camera-ready
% Figure~\ref{fig::fs2800} illustrates the high-level architecture of our DDR
% bus logic analyzer.

% \begin{figure}[h]
% \begin{center}
% \includegraphics[height=30mm]{./figs/fs2800.pdf}
% \end{center}
% \vspace*{-.5cm}
% \caption{High-level diagram of bus logic analyzer.}
% \label{fig::fs2800}
% \vspace*{-.25cm}
% \end{figure}

The traces gathered with the bus analyzer provide \emph{ground truth}
information about the rate of activations of a DRAM row and the memory
controller's behavior, including the logical addresses used to access DRAM.
We use these traces to characterize the ACT rates of different
Rowhammer instruction sequences from previous work and to construct a
sequence that has a near-optimal ACT rate on Skylake and Cascade Lake.

We found it difficult to use a high-level OS (e.g., Linux) for our
methodology for two reasons: (1) an OS introduces complex
virtual-to-physical address mappings that can change dynamically, and (2) an
OS's services introduce interfering traffic to a DIMM when testing.

Instead, our methodology boots the computer into the UEFI
mode~\cite{uefi2017uefi}.  In this mode, the virtual-to-physical address
map is linear and does not change across reboots. UEFI's simplicity and
lack of OS services eliminate any interfering DDR traffic from our traces.
However, it also increases the amount of engineering effort required to
implement our testing methodology because UEFI lacks many services commonly
found in a commodity OS. Therefore, we had to implement support for
multi-threading, hardware discovery~\cite{uefi2019acpi,dmtf2019smbios} and
performance counters.

\subsection{Performance Evaluation of Prior Instruction Sequences}
\label{sec::iseq}

Our results are based on experiments with six server-class DIMMs that one
cloud provider sourced from three different memory vendors, two DIMMs per
vendor. In alphabetical order, these vendors are: Hynix, Micron, and
Samsung. Although sourced from different vendors, the DIMMs' specs are
similar; they are registered ECC 32GB DDR4 (x4); the DIMMs from two of the
vendors have transfer rates of 2400 MT/s; and the third vendor's DIMMs have
rates of 2666 MT/s.  We found negligible differences in the performance of
an instruction sequence from one DIMM to another. For consistency, the
results presented in this section use the same DIMM.  One of the timing
parameters in the JEDEC specification is \emph{row cycle time} (tRC) -- the
minimum period of time between two back-to-back ACT commands.  The JEDEC
specification lists tables of minimum and maximum tRC values for different
types of DDR4 memory; these values depend on many memory characteristics,
such as speed, clock cycle, capacity, and so on.  Based on our DDR4
memory's characteristics, the JEDEC specification lists the \emph{minimum} value
of tRC as 47ns and does not specify a \emph{maximum} value (see Table 110
in~\cite{ddr4jedec}).

We measured tRC to be 46.7ns on all our hardware, corresponding to a
rate of 167.4 ACT commands between two consecutive REF commands issued by
the memory controller (i.e., one tREFI interval in JEDEC terminology).  We call
46.7ns the \emph{optimal} latency between two ACT commands, and 167.4
ACTs/tREFI the \emph{optimal} rate. All results presented are based on
experiments running on Skylake, although we ran many of these experiments
on Broadwell and Cascade Lake with similar results.  All servers use
motherboards with multiple CPU sockets.

% tREFI: 7812.5ns

%
% cloudflops uses two sequences: 
%   mov, mov, clflush, clflush
%   mov, mov, clflush, clflush, mfence
% 
% dedup-est-machine doesn't list assembly b/c their attack is JS-based
%    it corresponds to: mov, mov, cache-eviction, cache-eviction
%
% anotherflip: unclear what sequence they used
% 
% qiao2016new: they don't mount the attack, but present the following seq:
%   mov, mov, clflush, clflush, mfence
%      
% bhattacharya2016curious:
%  mov, mov, clflush, clflush
%
% jang2017sgx:
%  mov, mov, lfence, clflushopt, clflushopt, mfence
%
% aga2017good:
%  mov, mov, cache-eviction, cache-eviction
%

% Say how some do not say what they do.
% INTEL CAT (see nethammer), slow.

\begin{figure}[t]
\begin{lstlisting}[
    basicstyle=\footnotesize\ttfamily, %or \small or \footnotesize etc.
]
loop:
  movzx rax, BYTE PTR [rcx]
  movzx rax, BYTE PTR [rdx]
  clflush BYTE PTR [rcx]
  clflush BYTE PTR [rdx]
  mfence
  jmp loop
\end{lstlisting}
\vspace{-.35cm}
\caption{\ffsp Typical Rowhammer instruction sequence.}
\label{fig::iseq_typical}
\vspace{-.5cm}
\end{figure}

Previous work used a variety of different instruction sequences in a loop
to test for Rowhammer~\cite{kim2014disturbance, google2015projectzero,
  xiao2016cloudflops, bosman2016dedup-est-machina, qiao2016new,
  bhattacharya2016curious, gruss2016rowhammer-js, pessl2016drama,
  kaveh2016flip-feng-shui, veen2016drammer, lanteigne2016thirdio,
  jang2017sgx, aga2017good, gruss2017guide, gruss2018anotherflip,
  tatar2018defeating}. Some of these sequences use \emph{memory barriers}
to serialize the memory accesses on each iteration of the
loop~\cite{kim2014disturbance,xiao2016cloudflops,qiao2016new,jang2017sgx},
whereas others do
not~\cite{google2015projectzero,xiao2016cloudflops,bhattacharya2016curious,pessl2016drama,gruss2016rowhammer-js,veen2016drammer,aga2017good,tatar2018defeating}.
To bypass the cache hierarchy, some instruction sequences use an explicit
CPU flush instruction (e.g., \emph{clflush}), but not all do. Some use
cache collisions to flush the cache~\cite{lipp2018nethammer}; others
hypothesize that non-temporal load instructions~\cite{nontemporal} could
bypass the cache~\cite{veen2016drammer,gruss2017guide}. Another strategy we
encountered was choosing a pair of rows to hammer from a memory region
marked as uncached~\cite{gruss2017guide}. Finally, the x86-64 architecture
offers additional instructions for cache invalidation, such as
\emph{invd}~\cite{invd} and \emph{wbinvd}~\cite{wbinvd}.

The pseudo-code in Figure~\ref{fig::iseq_typical} describes a typical
sequence that issues two \emph{load} instructions, two \emph{clflush}
instructions, and a global memory barrier, all in one loop.  Several papers
on Rowhammer~\cite{kim2014disturbance, xiao2016cloudflops, qiao2016new},
including the original Rowhammer work~\cite{kim2014disturbance}, used this
sequence.

Faced with all these choices of possible instruction sequences, we
considered two questions: \textbf{(1)} \emph{Which prior instruction
  sequence maximizes the rate of ACT commands?} \textbf{(2)} \emph{How far
  from the optimal rate of ACT commands is the best instruction sequence?}

To answer these questions, we constructed 42 different
instruction sequences that let us experiment with:

\noindent
$\bullet$ All three types of fences available on x86-64 architectures:
\emph{mfence}~\cite{mfence}, \emph{lfence}~\cite{lfence}, and
\emph{sfence}~\cite{sfence}.

\noindent
$\bullet$ Both \emph{clflush} and \emph{clflushopt}~\cite{clflushopt}
commands (the latter is an optimized cache flushing command with weaker
ordering semantics than the former).

\noindent
$\bullet$ Marking as uncacheable the hammered memory pages' PTEs, which
eliminates the need to issue any CPU flush commands.

\noindent
$\bullet$ Both regular and non-temporal~\cite{nontemporal} memory accesses.

\noindent
$\bullet$ Using the \emph{invd}~\cite{invd} and \emph{wbinvd}~\cite{wbinvd}
commands to invalidate CPU caches.

\noindent
$\bullet$ Using a cache invalidation scheme based on cache line conflicts,
similar to the one used by Nethammer~\cite{lipp2018nethammer} and
ANVIL~\cite{anvil}.
% \item Hammering more than two adjacent rows that surround a victim row.
%   Previous papers~\cite{lanteigne2016thirdio} used a similar sequence to
%   increase the circuit noise around a victim. Our goal here is different --
%   we are investigating whether the rate of ACT commands increases when
%   accessing more rows from the same bank.

\begin{figure*}[t]
  \vspace{-.5cm}
  \begin{center}
    \includegraphics[width=\textwidth]{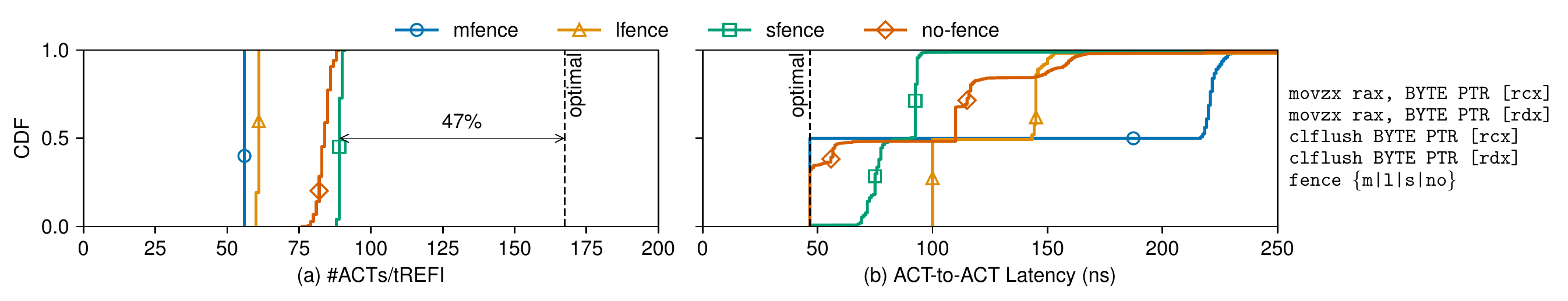}
  \end{center}
  \vspace*{-.5cm}
  \caption{\ffsp Performance of typical Rowhammer instruction sequences found
    in~\cite{kim2014disturbance,google2015projectzero,xiao2016cloudflops,bhattacharya2016curious,pessl2016drama,gruss2016rowhammer-js,veen2016drammer,qiao2016new,aga2017good,tatar2018defeating}. The
    left graph shows the CDF of the rate of ACT commands per tREFI; the right
    graph shows the CDF of the ACT-to-ACT latencies. The dotted black line
    corresponds to the optimal rate of ACT commands.}
  \label{fig::iseqs_0_21_22_1_samsung}
\end{figure*}

\begin{figure*}[t]
\vspace*{-.35cm}
  \begin{center}
\includegraphics[width=\textwidth]{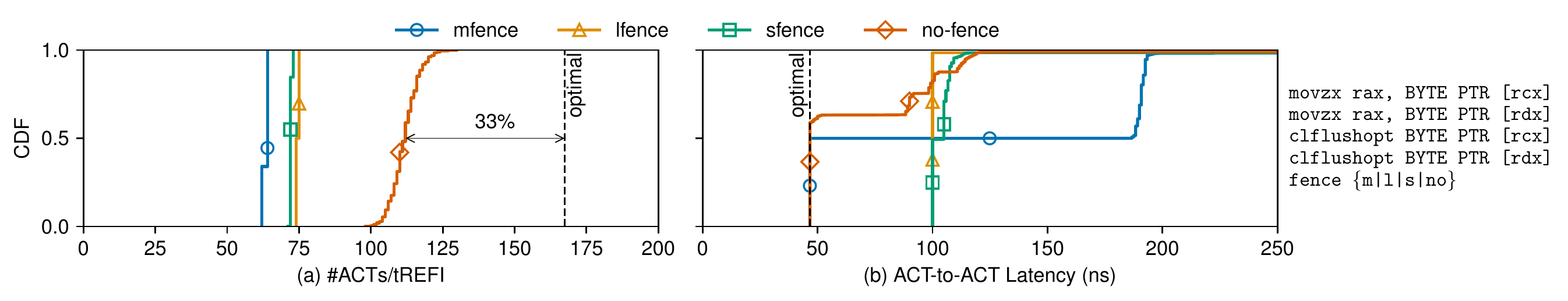}
\end{center}
\vspace*{-.5cm}
\caption{\ffsp Performance of  the same Rowhammer instruction sequences when using
  \emph{clflushopt} (rather than \emph{clflush}).}
\vspace*{-.5cm}
\label{fig::iseqs_35_37_38_36_samsung}
\end{figure*}

Figure~\ref{fig::iseqs_0_21_22_1_samsung} shows the performance of four
typical instruction sequences used in many previous Rowhammer attack
papers~\cite{kim2014disturbance, google2015projectzero, xiao2016cloudflops,
  bhattacharya2016curious, pessl2016drama, gruss2016rowhammer-js,
  veen2016drammer, qiao2016new, aga2017good, tatar2018defeating}. These
four sequences are identical except for the type of memory barrier they use
to serialize the reads and flushes on each loop iteration. The assembly
code is shown to the right of the graphs; the four sequences use
mfence~\cite{mfence}, lfence~\cite{lfence}, sfence~\cite{sfence}, and no
fence at all, respectively.

Figure~\ref{fig::iseqs_0_21_22_1_samsung}a shows the CDF of the rate of row
activations for each sequence as well as the optimal rate (with a dashed
line). Despite the popularity of these instruction sequences, which were
used by previous papers to mount Rowhammer attacks, we discovered that they
do not create worst-case ACT rates -- their rates of row activations are
\emph{47\% from optimal}. Even worse, using \emph{mfence}, a sequence used
by~\cite{kim2014disturbance, xiao2016cloudflops, qiao2016new}, leads to the
\emph{slowest} ACT rate of the four sequences. The most effective sequences
impose no ordering and use no fences or only store fences (sfence). A store
fence has no ordering effect because no sequences issue any stores.

Figure~\ref{fig::iseqs_0_21_22_1_samsung}b shows the CDF of the latencies
between two consecutive row activations for each instruction sequence.
Although the sequence using mfence has the slowest ACT rate, half of its
latencies are \emph{optimal}.  We examined its behavior closely and found
that the two reads inside the loop are always issued back-to-back at the
optimal latency (46.7ns).  However, the next ACT command is issued after a
long delay of over 220ns, which is caused by the \emph{mfence}.  The delay
explains why this sequence has a low rate of ACT commands
%(only 33\% of the optimal rate).
This bimodal delay between ACTs is clearly visible in
Figure~\ref{fig::iseqs_0_21_22_1_samsung}b.

These results illustrate the gap between a proof-of-concept Rowhammer
attack and the needs of a DRAM testing methodology. Although the
instruction sequences shown in Figure~\ref{fig::iseqs_0_21_22_1_samsung}
have been used in many papers to mount various successful Rowhammer attacks,
their ACT rates are \emph{far} from optimal. This suggests that
the DIMMs found vulnerable in previous work succumbed even to a low-rate
Rowhammer attack.  A DRAM testing methodology based on these instruction
sequences falls short of creating the \emph{worst-case} testing conditions
for DRAM, which is necessary to confidently determine whether a chip is
vulnerable to Rowhammer.

\vspace{.2cm}
\noindent
\textbf{Using \emph{clflushopt} improves ACT rates.}  With Skylake, Intel
introduced an optimized version of the cache flush instruction, called
\emph{clflushopt}, that has weaker ordering requirements than
\emph{clflush}~\cite{clflushopt,clflush_vs_clflushopt}.  Multiple
\emph{clflushopt} instructions to different cache lines can execute in
parallel. We performed a detailed analysis by implementing support for
performance counters in UEFI mode~\cite{uefi2017uefi}.  We found that using
\emph{clflushopt} in our instruction sequences takes only 3 micro-ops,
whereas \emph{clflush} takes 4.

Figure~\ref{fig::iseqs_35_37_38_36_samsung} shows the three instruction
sequences using memory barriers that have similar ACT rates; some have
slightly higher rates (the ones using \emph{sfence} and \emph{lfence}),
whereas others have slightly lower rates (the one using \emph{mfence}).
Although it is difficult to quantify how different types of barriers affect
the performance of the two cache line flush instructions, the high-level
finding remains the same: memory barriers are slow, and instruction
sequences using barriers have low ACT rates.

In contrast, the sequence that uses no memory barriers has a much higher
rate of ACT commands of 112 every tREFI, corresponding to 33\% from 
optimal. The lack of a memory barrier causes this instruction sequence to
have the \emph{highest} ACT rate overall. This finding is not intuitive --
the lack of memory barriers makes the CPU re-order memory accesses to
increase the likelihood that they will be served from the cache.  We
measured this sequence and found its cache hit rate to be 33\% (in
contrast, the sequence using \emph{mfence} has a 0\% cache hit rate).
Despite the CPU cache acting as a \emph{de facto} rate limiter to
Rowhammer attacks, the ACT rate of this instruction sequence is higher than
those using any type of barrier.

% \vspace{.2cm}
% \noindent
% \textbf{Alternative strategies are ineffective and have low ACT rates.}
% Appendix~\ref{appendix::other_optimizations} shows our results for other
% instruction sequences that were not evaluated by previous work.  We
% investigated the ACT rates for uncached PTEs, non-temporal memory accesses,
% full cache invalidations, cache collisions, and loads vs. stores.  We found
% that all these alternative instruction sequences have low ACT rates.

\begin{figure*}[t]
  \begin{center}
    \includegraphics[width=\textwidth]{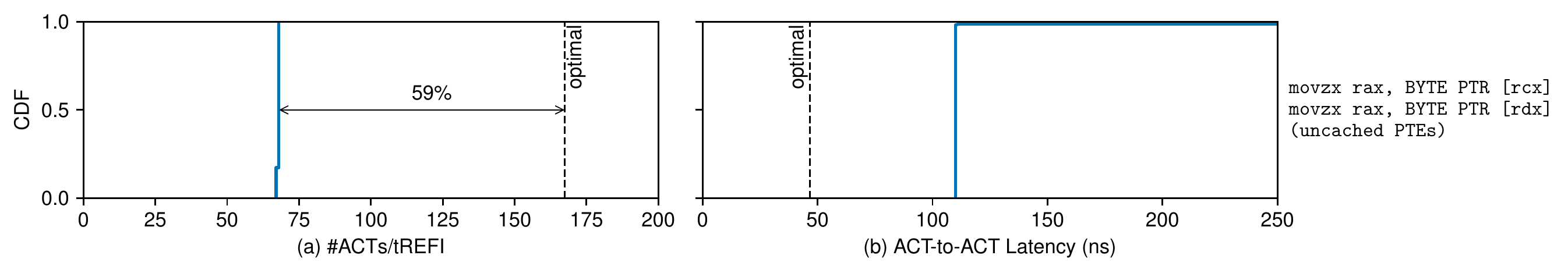}
  \end{center}
  \vspace*{-0.4cm}
\caption{\ffsp Performance of the Rowhammer instruction sequence that marks its memory
  pages uncacheable.}
\vspace{-.5cm}
\label{fig::iseq_7_samsung}
\end{figure*}

\begin{figure*}[t]
  \begin{center}
    \includegraphics[width=\textwidth]{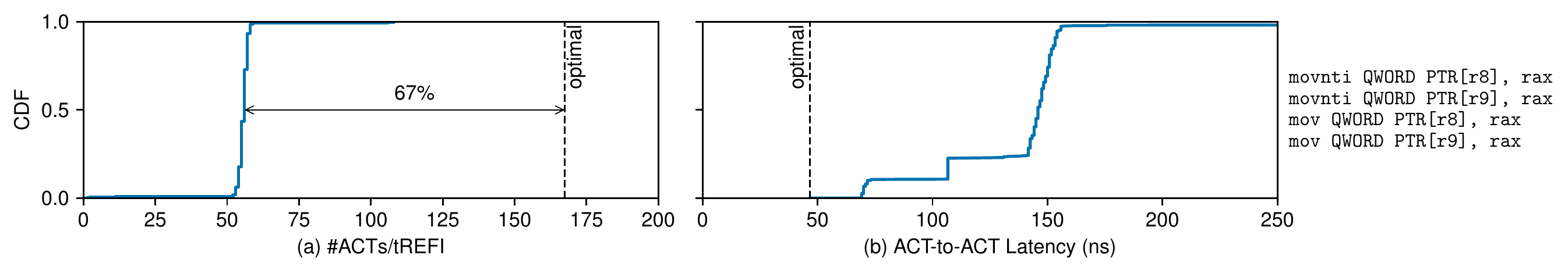}
  \end{center}
  \vspace*{-0.4cm}
\caption{\ffsp Performance of Rowhammer instruction sequence that uses a
  mix of non-temporal and regular memory accesses~\cite{gruss2017guide}.}
\vspace{-.5cm}
\label{fig::iseq_42_samsung}
\end{figure*}

\vspace{.2cm}
\noindent
\textbf{Uncached PTEs.} We experimented with an instruction sequence that
does not issue cache flushes, but instead marks its memory pages
uncacheable. The sequence has a low rate of ACT commands.

Figure~\ref{fig::iseq_7_samsung} shows the performance of an instruction
sequence that does not issue cache flushes, but instead marks its memory
pages uncacheable. Such a sequence has a low rate of ACT commands, and also
a very regular behavior: its ACT-to-ACT latencies are almost always 110ns
apart (in a small fraction of cases, this latency increases because the ACT is blocked
behind an ongoing refresh command). These results suggest that loading an
address from uncached memory has a fixed high cost, making this instruction
sequence have a low rate of ACT commands.

\vspace{.2cm}
\noindent
\textbf{Non-temporal memory accesses.} Intel offers five non-temporal store
instructions for different data types (e.g., integers, doublewords,
floating point, etc.) and one load instruction for double quadwords. These
instructions do not follow normal cache-coherence rules and fetch their
corresponding cache line from DRAM.

While experimenting with non-temporal instructions, we discovered that
accesses are \emph{cached}, and not served by DRAM.  According to Intel's
documentation, these accesses use a form of \emph{temporary internal
  buffers} that might prevent them from accessing DRAM, which can explain our
findings.

We also experimented with instruction sequences that combine non-temporal
and regular memory accesses. We expected that the different caching
semantics of these two types of memory accesses would flush the caches (or
internal buffers) in each loop iteration.  Previous work also proposed
mixing these two types of memory accesses for mounting
Rowhammer~\cite{gruss2017guide}. We found these instruction sequences to
be ineffective.  Figure~\ref{fig::iseq_42_samsung} shows the
performance of one such instruction sequence to be far from optimal (67\%
from optimal).

\vspace{.2cm}
\noindent
\textbf{Full cache invalidations, cache collisions, loads vs. stores.} We
experimented with replacing the cache line flush instructions with full cache
invalidation instructions: \emph{invd}~\cite{invd} and
\emph{wbinvd}~\cite{wbinvd}. We found that full cache invalidation
instructions are very expensive, making the instruction sequences have low
ACT rates. We also experimented with generating cache conflicts to evict cache
lines, but did not find higher ACT rates. Finally, we experimented
with replacing \emph{loads} with \emph{stores} in various instruction sequences
and found negligible performance differences.  For brevity, we omit showing
these results.

\vspace{.2cm}
\noindent
\textbf{Differences across memory vendors.}  While all results shown use
DRAM from one single vendor, we performed these experiments on DRAM from
all three vendors.  When using the same instruction sequence, we found no
significant differences in the ACT rates when using DRAM from different
vendors.

\vspace{.2cm}
\noindent
\textbf{Key Takeaways}

\noindent
$\bullet$ All previously proposed instruction sequences are sub-optimal for
two reasons: (1) memory barriers are expensive and thus reduce the ACT
rate, and (2) in the absence of memory barriers, the CPU re-orders memory
accesses to increase the cache hit rate and avoid accessing DRAM.

\noindent
$\bullet$ Using uncached PTEs or non-temporal instructions is ineffective,
leading to negligible changes to ACT rates when compared to the more common
Rowhammer instruction sequences.

\noindent
$\bullet$ The most effective instruction sequence proposed in previous
works uses two load, two \emph{clflushopt} instructions, and no memory
barriers at 33\% from the optimal rate
(Figure~\ref{fig::iseqs_35_37_38_36_samsung}).

\subsection{\emph{clflushopt} Alone Hammers Near-Optimally on Skylake and Cascade Lake}

To increase the rate of ACT commands, we experimented with new instruction
sequences that (1) do not use memory barriers, and (2) are less prone to
the effects of out-of-order execution. Our experiments revealed that a
\emph{cache line flush instruction results in a memory access}.

\begin{figure*}[t]
\vspace{-.5cm}
  \begin{center}
    \includegraphics[width=\textwidth]{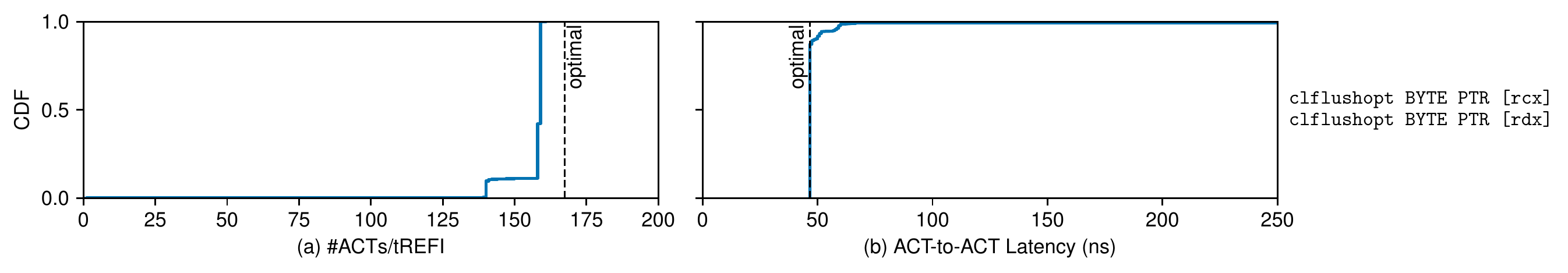}
  \end{center}
  \vspace*{-.5cm}
  \caption{\ffsp Performance of the near-optimal Rowhammer instruction sequence
    using only \emph{clflushopt}.}
  \label{fig::iseq_2162_samsung}
\end{figure*}

Figure~\ref{fig::iseq_2162_samsung} characterizes the rate of ACT commands
of a sequence consisting of two \emph{clflushopt} instructions in a loop;
(these results are from our experiments on the Skylake-based server, but
they are very similar to those performed on Cascade Lake).
Figure~\ref{fig::iseq_2162_samsung}b shows that over 87\% of ACTs are
issued \emph{at the optimal rate}, about 46.7ns apart from one another.
The remaining 13\% are separated by an additional 10-20ns due to conflicts
with ongoing refresh commands (REF).  When the memory controller issues a
REF, the bank remains inaccessible until the REF
finishes~\cite{chang2014improving}, and any ongoing ACT is blocked waiting
for the REF to finish. The REF-induced delay causes this instruction
sequence to issue 159 row activations for every tREFI, a rate we call
\emph{near-optimal}.

The microarchitectural side-effects of \emph{clflushopt} causes this
instruction sequence to issue row activations at a rate that is \emph{44\%
  higher} than the best previously known Rowhammer sequence (159 vs 110
ACTs/tREFI).  It is unlikely another sequence could improve this rate
because row activations will still conflict with REFs that block the bank.
Two \emph{clflushopt} instructions in a loop thus create the worst-case
DRAM ``hammering'' conditions on Skylake and Cascade Lake.

\vspace{.2cm}
\noindent
\textbf{Why does \emph{clflushopt} cause memory accesses?}  This
instruction sequence is highly surprising in the context of a Rowhammer
attack because it uses no explicit memory accesses. Instead, the memory
access (a DDR4 read operation) is a microarchitectural side-effect of the
CPU executing a cache line flush. It occurs \emph{only when}
the cache line is invalid. Issuing a cache line flush
instruction to a line present in the cache \emph{does not
  cause} any DDR read operations.

Our instruction sequence (Figure~\ref{fig::iseq_2162_samsung}) causes two
memory accesses for each loop iteration except for the first iteration. The
first loop iteration does not generate memory accesses when the lines are
in the cache. However, it invalidates the cache lines, causing all
subsequent iterations to generate two memory accesses.

\begin{figure*}[t]
  \vspace{-.5cm}
  \begin{center}
    \includegraphics[width=\textwidth]{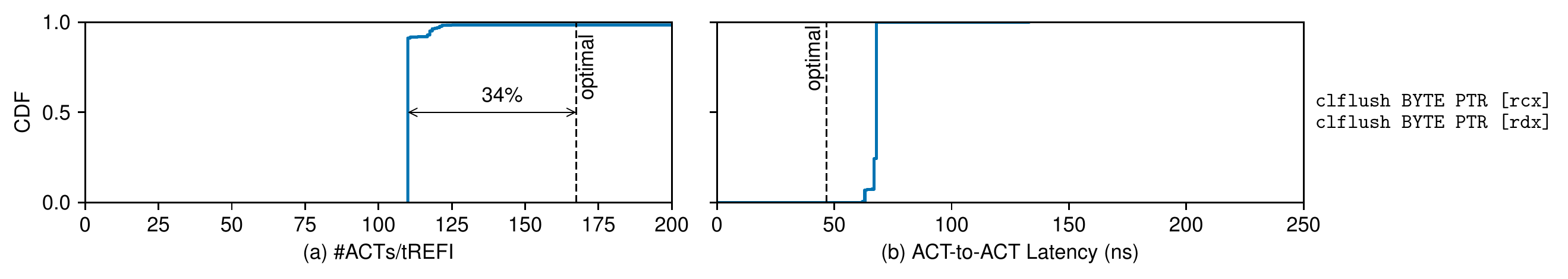}
  \end{center}
  \vspace*{-.35cm}
  \caption{\ffsp Performance of the Rowhammer instruction sequence using only
 \emph{clflush} instructions.}
  \vspace*{-.5cm}
\label{fig::iseq_2262_samsung}
\end{figure*}

According to Intel's specification~\cite{dirbits,mccalpin2018tacc}, systems
with multiple processors may maintain cache coherence state of each cache
line \emph{within the line itself in memory}. When executing
\emph{clflushopt} on an invalid cache line, the processor reads cache
directory state from DRAM to determine whether the line is present in other
processors' caches. We verified \emph{clflushopt}'s behavior on both
Cascade Lake and Skylake. We also show that \emph{clflush} behaves
similarly on both Cascade Lake and Skylake, but on Broadwell
\emph{clflush} results in no memory accesses.  We hypothesize that \emph{clflush} has
more system overhead than \emph{clflushopt} because it is subject to additional ordering
constraints~\cite{intel2019ordering}, leading to a reduced rate of DRAM row
activations.  Figure~\ref{fig::iseq_2262_samsung}
shows the performance of a sequence using two \emph{clflush} instructions
in a loop; it activates rows at a rate of 110 every tREFI, corresponding to
65.7\% of optimal.

\section{Step 2: Reverse Engineering Row Adjacency in Any DRAM Device}
\label{sec::fi}

No technique used in previous work is suitable for reverse engineering row
adjacency: some are not fine-grained and cannot determine adjacency at the
level of an individual row~\cite{schwarz2016drama,
  pessl2016drama,lee2017diva,jung2016crosshair}, whereas others do not
capture addresses internal to DRAM devices and thus can determine adjacency
only in the DDR4 bus address space~\cite{schwarz2016drama,pessl2016drama}.
The single previous technique that can overcome these limitations works
\emph{only if} the device succumbs to Rowhammer
attacks~\cite{schwarz2016drama, pessl2016drama, tatar2018defeating}.
Section~\ref{sec::methodology::challenges::adjacency} describes these
techniques and their trade-offs in depth.

In an attempt to guarantee Rowhammer failures on our DRAM devices, we
experimented with lowering the refresh rates of our servers. A low refresh
rate ensures that an attack sends a higher number of ACT commands to a
victim row before the row can refresh. This increases the attack's
likelihood of success. Unfortunately, our experiments were unsuccessful.
Recent hardware makes it increasingly difficult to set the refresh rates
sufficiently low to successfully mount a Rowhammer attack.  Older
generation BIOSes running on DDR3-equipped hardware can set refresh rates
up to 12x lower than normal; such low refresh rates make DDR3 devices
succumb to Rowhammer attacks.

Modern BIOSes for DDR4 hardware restrict lowering the rate to only $\sim$3.5x.
Unfortunately, this refresh rate is not sufficiently low to \emph{guarantee}
Rowhammer failures on our servers. We also confirmed this is not a GUI
restriction: we examined the BIOS source code and found that the refresh
interval configuration register cannot hold a value larger than one
corresponding to a refresh rate of 3.5x lower.

\begin{figure}[t]
  \begin{center}
    \includegraphics[height=21mm]{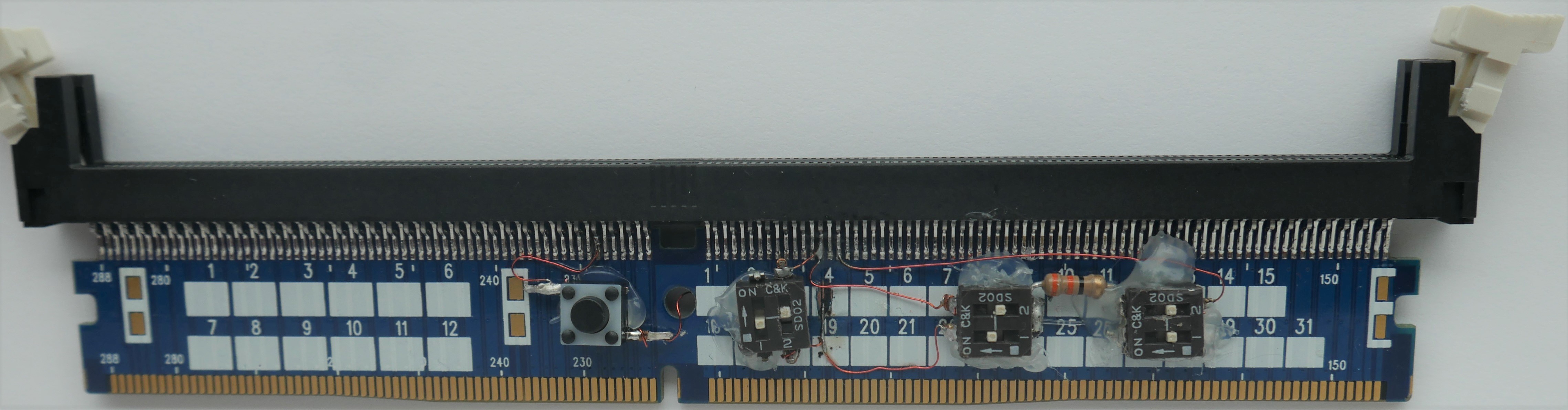}
  \end{center}
  \vspace*{-.3cm}
  \caption{\ffsp Fault injector. When pressed, the button drives A14 to
    low. Two DIP switches form a 3-way switch to flip the $ALERT_n$ signal.
    The third DIP switch is a spare.}
  \label{fig::fi_ver2}
  \vspace*{-.4cm}
\end{figure}

\begin{figure}[t]
  \begin{center}
    \includegraphics[width=.25\textwidth]{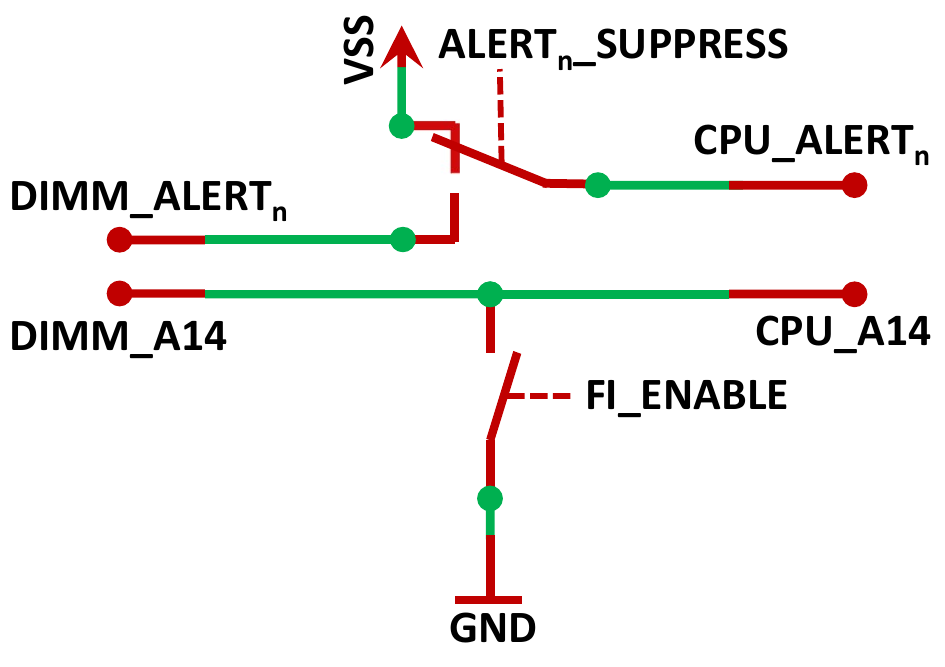}
  \end{center}
  \vspace*{-.3cm}
  \caption{\ffsp Fault injector schematic.}
  \label{fig::fi_schematic}
    \vspace*{-.6cm}
\end{figure}
%Figure~\ref{fig::fi_schematic} shows the schematic of our fault injector. 

\subsection{DDR4 Fault Injector}

Because modern BIOSes restrict lowering the DRAM refresh rate, we used a
different approach: we designed a DDR4 fault injector that blocks REFs sent
by a memory controller to an individual DIMM. Our fault injector
manipulates electrical signals and drives them from low to high, and
vice-versa.  Manipulating the DDR bus's electrical signals effectively
changes one DDR command into another. This insight was inspired by previous
work that used a custom-made shunt probe to induce faults in DRAM data and
thus reverse engineered the DRAM controller's ECC
scheme~\cite{cojocar2019ecc}.  Figure~\ref{fig::fi_ver2} shows our fault
injector, and Figure~\ref{fig::fi_schematic} shows its schematic.

\vspace{.2cm}
\noindent
\textbf{Side-effects.}  Manipulating electrical signals to change DDR4
commands introduces side-effects. For example, changing a signal known as
$\overline{\mbox{ACT}}$ makes all DDR4 commands decode as row activate
(ACT) commands.  In this case, a DIMM becomes inaccessible because it
receives only ACTs no matter what command the memory controller is issuing.
Table~\ref{tbl::ddr4_cmds} (reproduced from Wikipedia~\cite{wiki2019DDR4})
shows the encoding of DDR4 commands.

\begin{table}[t]
%\vspace{-.3cm}
  \begin{center}
    \includegraphics[width=.5\textwidth]{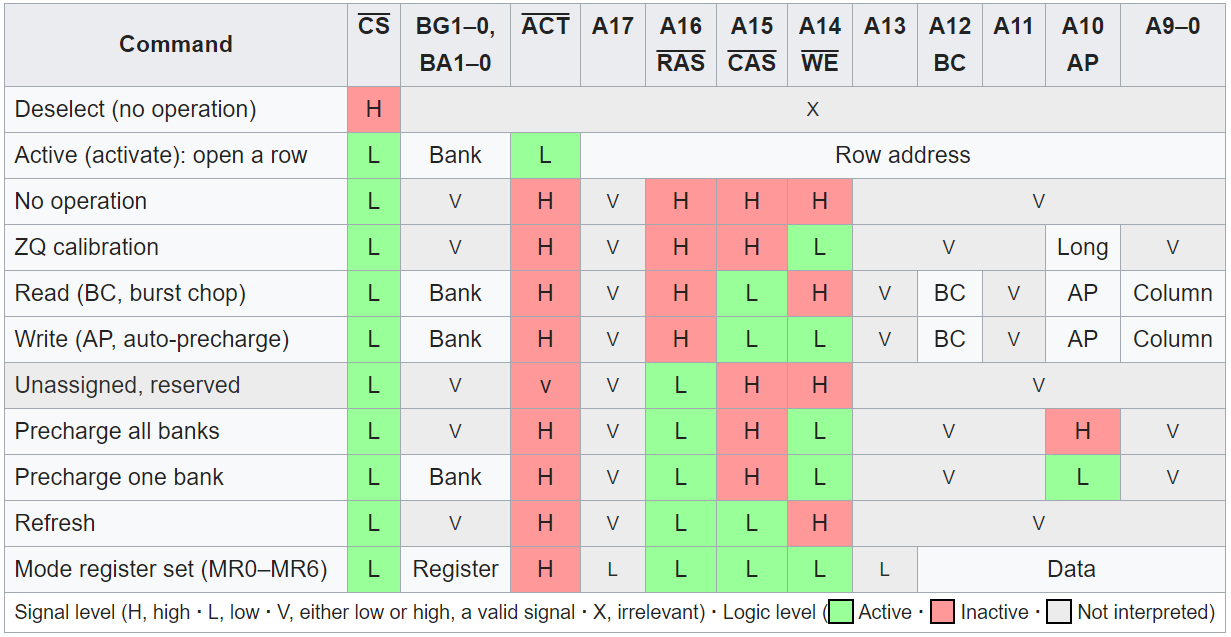}
  \end{center}
  \vspace*{-.3cm}
  \caption{\ffsp DDR4 command encoding~\cite{wiki2019DDR4}.}
  \vspace*{-.5cm}
  \label{tbl::ddr4_cmds}
\end{table}

Instead, we need to control these side-effects to leave the DIMM in a
responsive state; otherwise, we cannot mount a Rowhammer attack. The DIMM
must continue to receive row activates, row reads (or writes), and row
precharges.

\vspace{.2cm}
\noindent
\textbf{Overcoming the side-effects.}  Our fault injector changes the A14
signal from high to low and turns REF commands into a different DDR
command, known as mode register (MR0)~\cite{ddr4jedec} with a null payload.
Although this new command affects the DIMM's configuration, the DIMM
continues to serve all incoming commands. We designed our fault
injector to trigger memory recalibration and thus reset the DIMM's
configuration back to its original settings.

Manipulating the A14 signal has an additional side-effect: it changes a
read into a write command (see~\cite{wiki2019DDR4}). To overcome this
side-effect, our Rowhammer attack instruction sequence uses stores rather
than loads. Fortunately, manipulating the A14 signal does not affect the
row activations and precharges needed to mount a Rowhammer attack.

\vspace{.2cm}
\noindent
\textbf{Memory recalibration.} The memory controller performs memory
recalibration upon detecting an error. One such error is a parity check
failure for DDR4 signals.  On an incoming command, the DIMM checks parity,
and, if the check fails, it alerts the memory controller through a reverse
signal called $ALERT_n$.  Upon receiving the alert, the memory controller
sends a sequence of DDR recalibration commands to the DIMM.

We designed our fault injector to also recalibrate memory, but \emph{only
  when the Rowhammer attack completes}. This restores the DIMM to its
original configuration and lets us inspect the location of the bit flips
that could reverse engineer row adjacency in the DRAM device.  Memory
recalibration cannot occur \emph{during} an ongoing Rowhammer attack
because it creates interference.

To recalibrate memory, our fault injector also manipulates the $ALERT_n$
signal. During an ongoing Rowhammer attack, it suppresses the $ALERT_n$
signal, thus preventing the memory controller from receiving any alerts.
Once the attack completes, the fault injector re-enables $ALERT_n$ while
continuing to manipulate A14 to ensure that parity checks continue to fail.
These alerts are now received by the memory controller, which, in turn,
recalibrates the DIMM.

\begin{figure}[t]
  \begin{center}
    \includegraphics[height=48mm]{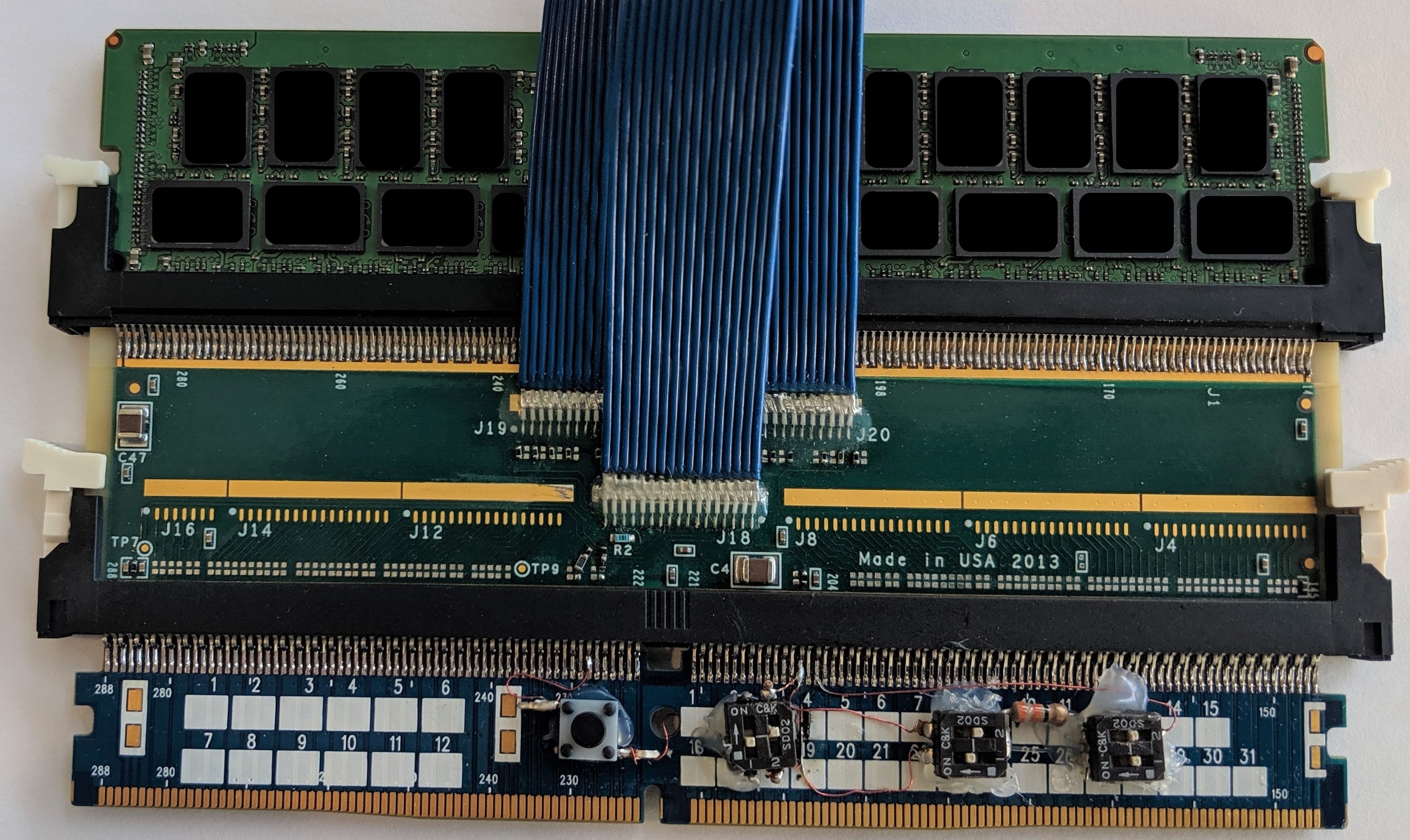}
  \end{center}
  \vspace*{-.25cm}
  \caption{\ffsp Our hardware stack from the bottom up: fault injector, bus analyzer's
    interposer, and DDR4 DIMM. Markings on the DRAM chips have been
    redacted.}
  \vspace*{-.5cm}
  \label{fig::fi_stacked}
\end{figure}

\vspace{.2cm}
\noindent
\textbf{Methodology for injecting DDR4 faults.}
Figure~\ref{fig::fi_stacked} shows our hardware stack: the fault injector,
the bus analyzer's interposer, and the DDR4 DIMM. We used an eight-step
operational plan to inject faults and mount Rowhammer to induce bit flips
capable of reverse engineering row adjacency:

\vspace{.1cm}
\noindent
1. Boot server with DDR parity check enabled and ECC disabled.

\noindent
2. Suppress $ALERT_n$ signal with DIP switches.

\noindent
3. Begin Rowhammering the target DIMM.

\noindent
4. Inject a fault in the A14 signal by pressing the button switch for a
fixed time interval. During this time, the DIMM receives no REFs, and the
memory controller receives no alerts.

\noindent
5. Stop Rowhammering the target DIMM.

\noindent
6. Re-connect $ALERT_n$ signal with DIP switches.

\noindent
7. Inject a fault in the A14 signal by tapping the button. The memory controller
  receives alerts from the DIMM and starts recalibrating the DIMM.

\noindent
8. Inspect the number and spatial distribution of bit flips.

\subsection{Row Adjacency in DRAM Devices}
\label{sec::re_dram}

We used the fault injector to reverse engineer the physical row adjacency
of DRAM devices mounted on DDR4 DIMMs.  We mounted Rowhammer attacks and
measured the density of bit flips across each row within a bank. This
technique correlates each row's density of bit flips with
adjacency~\cite{schwarz2016drama, pessl2016drama, tatar2018defeating}.
Invariably, when hammering one row for 15 seconds without refreshes, a
small number of rows flip bits at a much higher rate than all others. This
indicates that these highly affected rows are \emph{physically adjacent} to
the hammered one.

We then posed the following questions:

\emph{1. Do logical addresses map linearly to internal DRAM
  addresses?}  A linear map makes it easier to mount Rowhammer because an
attacker need not reverse engineer it.  Previous work showed that the map
from physical to logical addresses is not linear and discussed how
non-linearity can render many of the Rowhammer defenses much less effective
than initially thought~\cite{tatar2018defeating}.

\emph{2. Does the position of a bit within a word influence its likelihood
  of being flipped?} Such results would shed light on whether some words
(or some bits within a single word) are more susceptible to Rowhammer
attacks than others. For example, most page table entries have a format in
which low-order bits control access to a page; should the low-order bits be
more susceptible than high-order, an attack changing the access control
policy to a page would be more likely to succeed.

\emph{3. How do data patterns affect the susceptibility of bits
  being flipped?} We examined the direction in which bits flip (0-to-1 or
1-to-0). The memory controllers in datacenter servers are routinely
configured to \emph{scramble} data in DRAM by xor-ing it with a known,
random data pattern~\cite{mosalikanti2011ddr,intel2018datasheet}.  This
means that the proportion of 0s to 1s in DRAM is 50-50.

\emph{4. Do DIMMs sourced from different vendors have different
  characteristics?} We examined whether or not the map and the rate at
which bits flip are consistent across DIMMs from different vendors.

\vspace{.2cm}
\noindent
\textbf{Methodology.}  We performed all experiments by suppressing REFs for
15 seconds at room temperature. We disabled data scrambling and wrote a
specific data pattern across the entire bank except for the hammered row.
We wrote the complement of the data pattern in the hammered row, a strategy
used by previous work~\cite{ji2019pinpointing}.  We experimented with four
different data patterns that vary the locations and ratios of bits set to 1
vs. bits set to 0. The four patterns are: all 1s, 0xB6DB6DB...
(corresponding to two-thirds 1s), 0x492492... (corresponding to one-third
1s), and all 0s.  Unless marked otherwise, the results we present use a
pattern of all 1s.

When testing DRAM, a double-sided Rowhammer attack (i.e., two aggressor
rows) is better than single-sided (i.e., one aggressor row).  However, when
injecting faults, both types of Rowhammer attack flip bits because the
DIMM does not refresh for 15 seconds.  When reverse engineering row
adjacency, single-sided Rowhammer is simpler because the adjacency of a
flipped bit is unambiguous -- it is due to a single aggressor. Reverse
engineering row adjacency with double-sided Rowhammer leads to an
attribution challenge -- is the flipped bit adjacent to aggressor \#1 or
aggressor \#2? The results in this section are based on using a single
aggressor row, similar to mounting a single-sided Rowhammer attack.

Our results show no differences between the DIMMs from the same memory
vendor.  Most results are similar across DIMMs supplied by different
vendors; in these cases, we present the results from a single vendor
(referred to as vendor \#1). However, we note different vendors in the text
when results differ across vendors.

We verified that our fault injector suppresses refreshes on all three Intel
server architectures. We also reverse engineered portions of the row
adjacency maps of three DIMMs (one from each vendor) on both Broadwell and
Skylake and checked that the results are identical on both platforms. On
Cascade Lake, we reverse engineered only one DIMM, with identical results to Broadwell and Skylake.
The data shown in the remainder of this section was gathered on the Skylake
platform.

\begin{figure}[t]
  \vspace{-.35cm}
  \begin{center}
    \includegraphics[height=36mm]{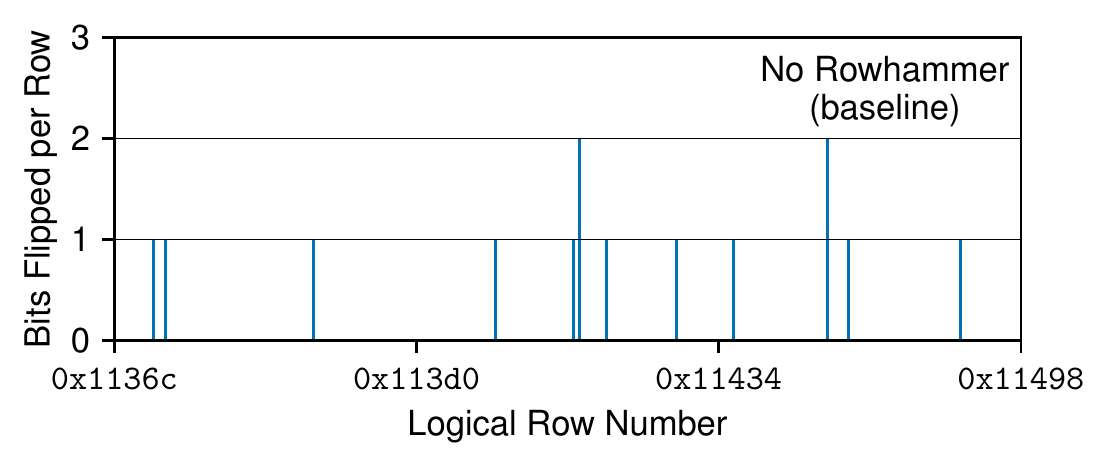}
  \end{center}
  \vspace*{-.5cm}
  \caption{\ffsp Number of bit flips per row when suppressing REFs for
    15 sec in the absence of Rowhammer, for a representative set of rows.}
  \vspace*{-.5cm}
  \label{fig::re_norh_micron_15s}
\end{figure}

\vspace{.2cm}
\noindent
\textbf{Results.}  We started with a baseline experiment in which we
suppressed REFs for 15 seconds without any additional memory workload. Our
goal was to determine the rate of bit flips due \emph{solely} to
suppressing REFs. These bit flips are not correlated with physical row
adjacency. If we observed many such bit flips, combining suppressing REFs
with Rowhammer attacks would make it difficult to attribute responsibility
for the bit flips.

Fortunately, Figure~\ref{fig::re_norh_micron_15s} shows that the number of
flipped bits is low. Even after suppressing REFs for 15 seconds, the
majority of rows show no failures. Ten rows have a single flipped bit (a
row failure rate of 3.3\%), and two have two flipped bits (a row failure
rate of 0.7\%). No row has more than two flipped bits. The low number of
failures demonstrates that our DRAM is resilient when not refreshed \emph{in the
  absence of Rowhammer attacks}.

\begin{table}[t]
%  \vspace{-.3cm}
  {\footnotesize
    \begin{tabular}{@{}c@{\hspace{.3cm}}c@{\hspace{.2cm}}c@{\hspace{.2cm}}c@{}}
      \toprule
      Aggressor&Victim \#1&Victim \#2&Victim \#3\\
      \midrule
    \textbf{0x11408}&0x11409 (75.7\%)&\textbf{0x11407 (40.0\%)}&\textbf{0x11417 (36.7\%)}\\
    0x11409&0x11408 (76.6\%)&0x1140A (76.5\%)\\
    0x1140A&0x11409 (75.3\%)&0x1140B (74.2\%)\\
    0x1140B&0x1140C (80.3\%)&0x1140A (79.5\%)\\
    0x1140C&0x1140B (77.0\%)&0x1140D (76.5\%)\\
    0x1140D&0x1140C (76.6\%)&0x1140E (75.9\%)\\
    0x1140E&0x1140D (77.5\%)&0x1140F (76.6\%)\\
    \textbf{0x1140F}&0x1140E (77.5\%)&\textbf{0x11410 (39.9\%)}&\textbf{0x11400 (37.7\%)}\\
    \textbf{0x11410}&0x11411 (77.7\%)&\textbf{0x1140F (40.5\%)}&\textbf{0x1141F (37.7\%)}\\
    0x11411&0x11412 (77.0\%)&0x11410 (76.7\%)\\
    0x11412&0x11411 (78.1\%)&0x11413 (77.2\%)\\
    0x11413&0x11414 (77.1\%)&0x11412 (76.4\%)\\
    0x11414&0x11413 (74.7\%)&0x11415 (74.0\%)\\
    0x11415&0x11414 (77.8\%)&0x11416 (77.4\%)\\
    0x11416&0x11415 (79.1\%)&0x11417 (78.3\%)\\
    \textbf{0x11417}&0x11416 (75.8\%)&\textbf{0x11418 (39.4\%)}&\textbf{0x11408 (36.8\%)}\\
    \bottomrule
  \end{tabular}}
  \setlength\tabcolsep{6pt} % default value: 6pt
%  \vspace*{-.25cm}
  \caption{\ffsp Adjacency for 16  rows consecutive in the logical address
    space.}
  \vspace{-.25cm}
  \label{tbl::row_adjacency}
\end{table}

\begin{figure*}[t!]
%  \vspace{-.1cm}
  \begin{center}
    \begin{subfigure}[b]{0.5\textwidth}
      \begin{center}
        \includegraphics[height=35mm]{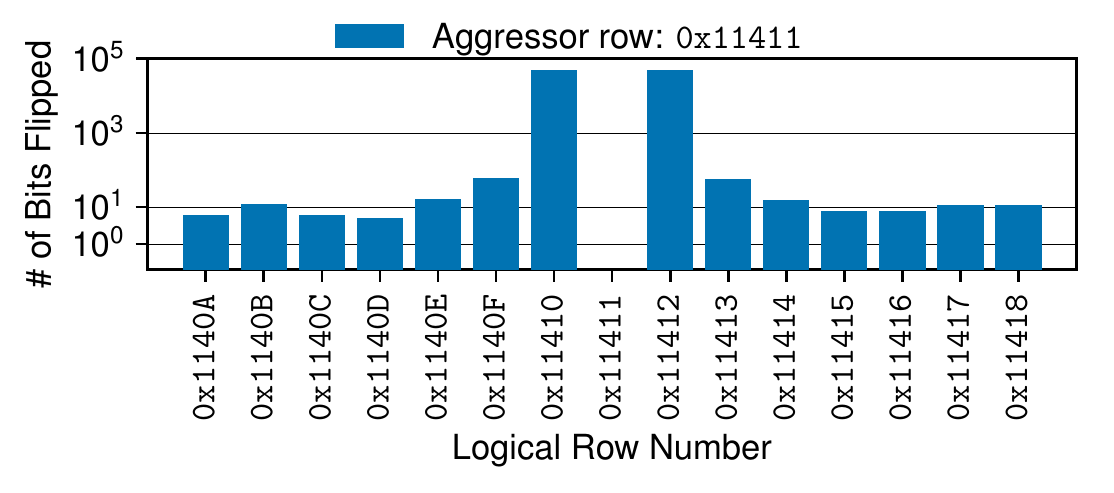}
        \vspace{-.35cm}
        \caption{Victim rows: 0x11410 and 0x11412.}
      \end{center}
    \end{subfigure}%
    ~
    \begin{subfigure}[b]{0.5\textwidth}
      \begin{center}
        \includegraphics[height=35mm]{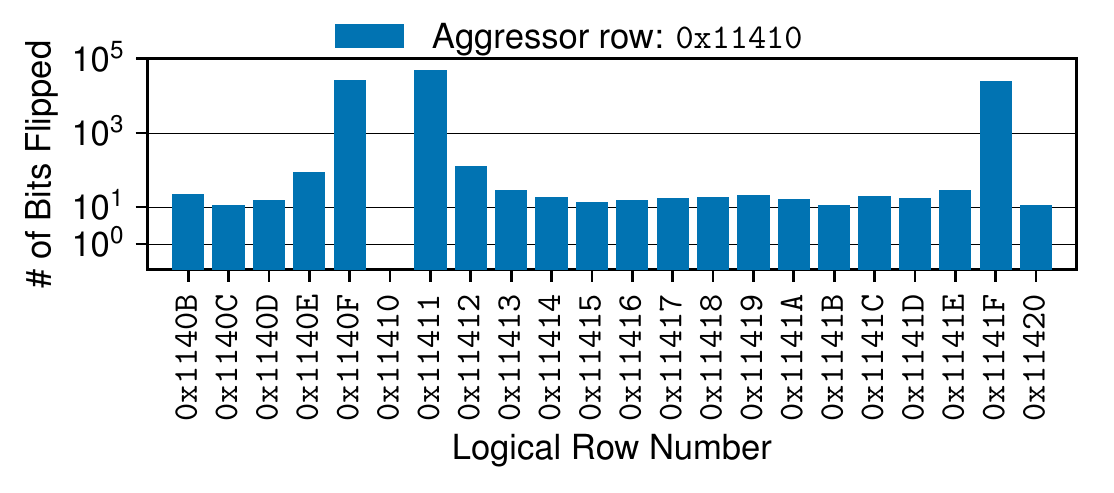}
        \vspace{-.35cm}
        \caption{Victim rows: 0x11411, 0x1140F, and 0x1141F.}
      \end{center}
    \end{subfigure}
  \end{center}
  \vspace*{-.35cm}
  \caption{\ffsp No. of bit flips on rows with neighboring logical row
    numbers (y-axis is logarithmic). A row has 65,536 bits.}
  \vspace*{-.1cm}
  \label{fig::re_micron_rows_0x11411_0x11410}
\end{figure*}

\vspace{.2cm}
\noindent
\textbf{Logical rows do not always map linearly.}
Figure~\ref{fig::re_micron_rows_0x11411_0x11410}a shows the number of
flipped bits per row when the aggressor was row 0x11411 (logical address).
The results suggest a linear map because most bit flips occur on rows
0x11410 and 0x11412 (the y-axis is logarithmic).  Row 0x11410 has 50,274
flipped bits (out of 65536, or 76.7\%), and row 0x11412 has 50,489
(77.0\%). All other rows have significantly fewer flipped bits.

However, the map is not always linear.
Figure~\ref{fig::re_micron_rows_0x11411_0x11410}b shows the results when
the aggressor row address was 0x11410. While victim row 0x11411 is adjacent
(with over 77\% of its flipped bits), victim row 0x1140F, although adjacent
in the logical address space, has only 26,566 flipped bits corresponding to
40.5\% of its bits.  Instead, a third victim row (0x1141F) has 24,680 of
its flipped bits (37.7\%). These results indicate that the aggressor row
0x11410 is adjacent to \emph{half} of victim rows 0x1140F and 0x1141F.  For
brevity, we say \emph{a victim has whole-row (or half-row) adjacency} to
refer to the adjacency of the victim to the aggressor row.

Table~\ref{tbl::row_adjacency} presents the row adjacency map inside the
DRAM device for 16 consecutive rows in the logical address space.  We
conducted 16 experiments in which a different row acts as the aggressor,
and list the top two or three victim rows sorted by the fraction of their
bits that flip (shown in parentheses); all remaining victims have fewer
than 1\% of their bits flip.  The data shows \emph{bimodal} behavior: many
rows map linearly (whole-row adjacency), but some have each of their halves
mapped differently (half-row adjacency).  Half-row adjacency in shown in bold in
Table~\ref{tbl::row_adjacency}.

When examining different portions of the adjacency map, we found that
half-rows occur frequently but lack a specific pattern.
Table~\ref{tbl::row_adjacency_bank_edge} shows the adjacency map for the
first rows of the bank. Row 0 is half-adjacent to row F, but the remaining
half is not adjacent to any other row in the bank. Half of row 0 is located
either next to a \emph{spare row}~\cite{hou2016spare} or on the physical
edge of the DRAM array. Also, some rows shown in
Table~\ref{tbl::row_adjacency_bank_edge} have an adjacency pattern
different from all others. For example, row 7 is half adjacent to rows 8
and 0x07f8.

\begin{table}[t]
  \begin{center}
  {\footnotesize
    \begin{tabular}{@{}c@{\hspace{.3cm}}c@{\hspace{.3cm}}c@{\hspace{.3cm}}c@{}}
      \toprule
      Row&\multicolumn{3}{c}{Adjacent Rows}\\
      \midrule
      \textbf{0x0000}&0x0001 (W)& \textbf{0x000F (H)}&\mbox{spare row/bank edge (?)}\\
      0x0001&0x0000 (W)& 0x0002 (W)\\
      0x0002&0x0001 (W)& 0x0003 (W)\\
      0x0003&0x0002 (W)& 0x0004 (W)\\
      0x0004&0x0003 (W)& 0x0005 (W)\\
      0x0005&0x0004 (W)& 0x0006 (W)\\
      0x0006&0x0005 (W)& 0x0007 (W)\\
      \textbf{0x0007}&0x0006 (W)&\textbf{0x0008 (H)} & \textbf{0x07F8 (H)}\\
      \textbf{0x0008}&0x0009 (W)& \textbf{0x0007 (H)}& \textbf{0x0017 (H)}\\
      \bottomrule
    \end{tabular}}
  \setlength\tabcolsep{6pt} % default value: 6pt
 % \vspace*{-.25cm}
  \caption{\ffsp Adjacency for the first rows in the bank. (W) represents
    half-row adjacency; (H) half-row.}
  \vspace{-.85cm}
  \label{tbl::row_adjacency_bank_edge}
\end{center}
\end{table}

\begin{figure*}[t]
  \vspace{-.35cm}
  \begin{center}
    \begin{subfigure}[b]{0.5\textwidth}
      \begin{center}
        \includegraphics[height=35mm]{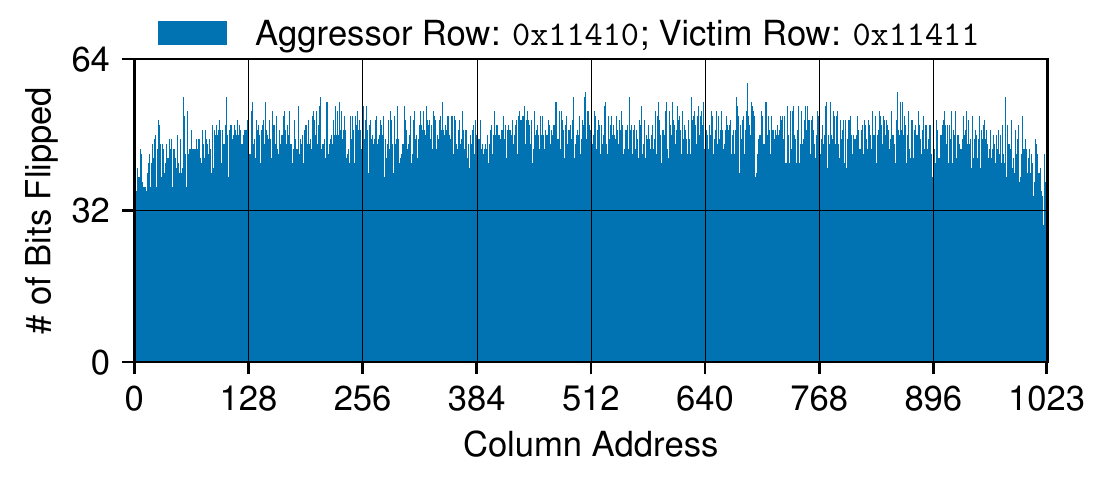}
        \vspace{-.35cm}
        \caption{Victim has whole-row adjacency.}
      \end{center}
    \end{subfigure}%
    ~
    \begin{subfigure}[b]{0.5\textwidth}
      \begin{center}
        \includegraphics[height=35mm]{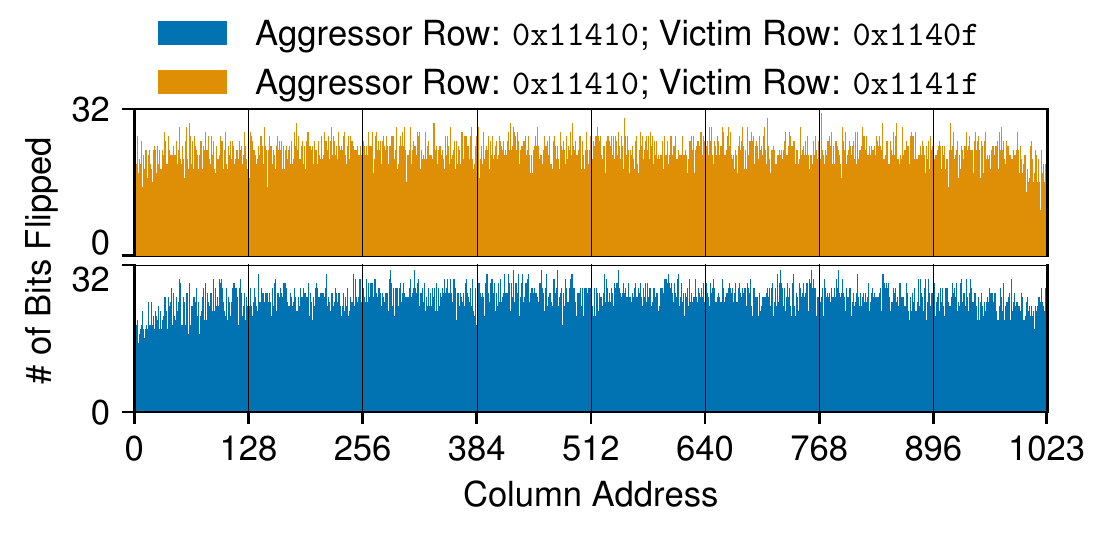}
        \vspace{-.35cm}
        \caption{Two victim rows, each with half-row adjacency.}
      \end{center}
    \end{subfigure}
  \end{center}
  \vspace*{-.35cm}
  \caption{\ffsp No. of bit flips in each word of a victim row (the
    column address specifies the word). A word has 64 bits.}
  \label{fig::re_rh_micron_row_col_index}
\end{figure*}

\begin{figure*}[t]
  \vspace{-.35cm}
  \begin{center}
    \begin{subfigure}[b]{0.5\textwidth}
      \begin{center}
        \includegraphics[height=35mm]{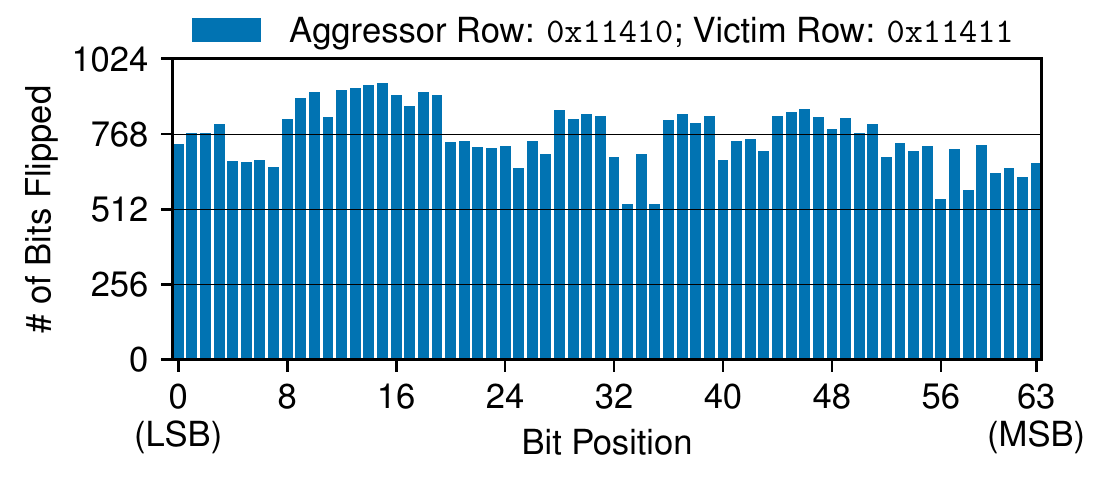}
        \vspace{-.35cm}
        \caption{Victim has whole-row adjacency.}
      \end{center}
    \end{subfigure}%
    ~
    \begin{subfigure}[b]{0.5\textwidth}
      \begin{center}
        \includegraphics[height=35mm]{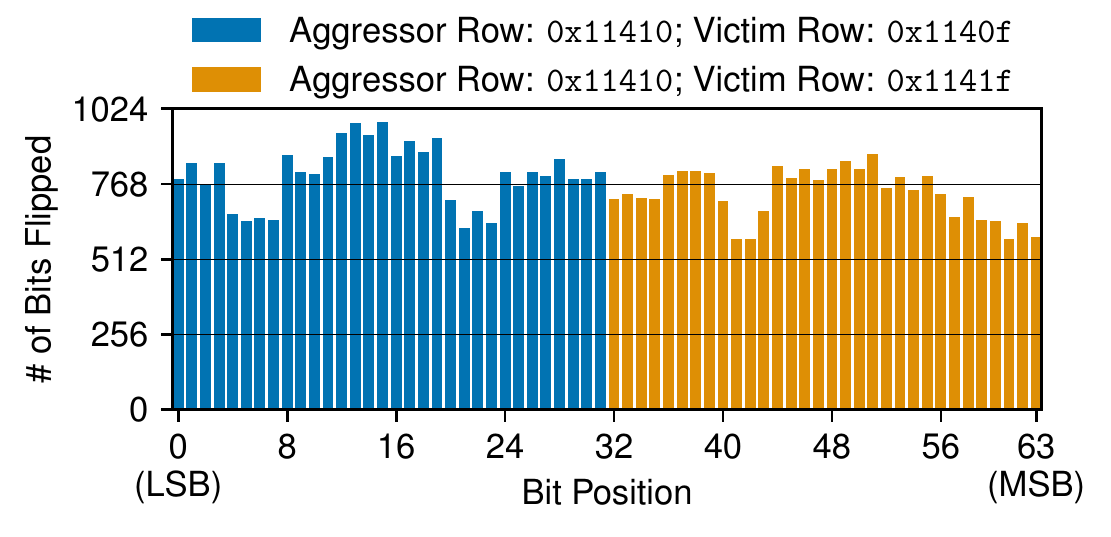}
        \vspace{-.35cm}
        \caption{Two victim rows, each with half-row adjacency.}
      \end{center}
    \end{subfigure}
  \end{center}
  \vspace*{-.4cm}
  \caption{\ffsp No. of bit flips in a representative victim row by their bit index
    positions within a memory word.}
  \vspace{-.45cm}
  \label{fig::re_rh_micron_row_bit_index}
\end{figure*}

\vspace{.2cm}
\noindent
\textbf{Fewer bits flip in half-rows than in whole-rows.}  We
characterize whether a bit's position influences its likelihood to be
flipped. We start by grouping bit flips in a victim row by their column
addresses (a column address specifies a word). A row has 1024 columns, and
a column contains a 64-bit word.

Figure~\ref{fig::re_rh_micron_row_col_index}a shows data from a whole-row
victim. Although some words have more bit flips than others, this variation
is relatively low: more than 95\% of all words have 40 to 60 bit flips, and
a word has 47.6 bit flips on average.  This result suggests that an
aggressor row affects all words in a victim whole-row more or less equally.

Half-row victims have \emph{half} the number of bit flips per word (the
remaining non-adjacent half is safe).
Figure~\ref{fig::re_rh_micron_row_col_index}b shows the number of bit flips
per word found in two half-row victims.  Each word has 25 and 23 bit flips,
respectively, on average as opposed to 47.6 bit flips for words located in a whole-row. This
result suggests that an aggressor row affects \emph{fewer} bits per word in
victim half-rows than victim whole-rows.

\vspace{.2cm}
\noindent
\textbf{All bits are equally susceptible for whole-row, but not for
  half-row, adjacency.}  We further investigated whether the position of a
bit in a word affects its likelihood of being flipped? For this, we
re-plotted the data from Figure~\ref{fig::re_rh_micron_row_col_index} by
grouping bits by their \emph{bit position} rather than their column
address. A row has 1024 bits in each bit position. Bit positions are
indexed 0 to 63 from the least to the most significant bit.

Figure~\ref{fig::re_rh_micron_row_bit_index}a shows the results of a
whole-row victim. As before, an aggressor row affects all bits in a victim
whole-row equally.  Figure~\ref{fig::re_rh_micron_row_bit_index}b shows the
results for two half-row victims. Surprisingly, the sets of bit positions
are \emph{disjoint}. One of the half-row victims has bit flips in positions 0
through 31 \emph{only}, whereas the other has bit flips in positions 32
through 63.

These results indicate that the position of a bit determines its likelihood
of being flipped \emph{in half-rows only}. In a victim half-row, either the
most significant (63-32) or the least significant (31-0) bits are flipped,
depending on memory geometry.  For little endian systems (such as ours),
the region containing bit flips will inversely map to the most or least significant bits,
respectively. For example, in big endian systems, the map will be direct: words in row
0x1140F will have their least significant bits flipped, while words in row
0x1141F, their most significant. All DIMMs from all three hardware vendors
exhibited this behavior.

\vspace{.2cm}
\noindent
\textbf{Most, but not all, bits flip from 1 to 0.}  Electromagnetic
coupling (considered to be a main reason for
Rowhammer~\cite{kim2014disturbance}) drains capacitors faster than normal.
Memory encodings can represent a charged capacitor as either a 1 or a 0,
making the data pattern another factor in a bit's susceptibility to be
flipped.  Cells that encode data value 1 as a charge are called
\emph{true-cells}, while \emph{anti-cells} encode data value 0 as a
charge~\cite{kira2018cells}.

\begin{figure}[t]
  \vspace{-.35cm}
  \begin{center}
    \includegraphics[height=36mm]{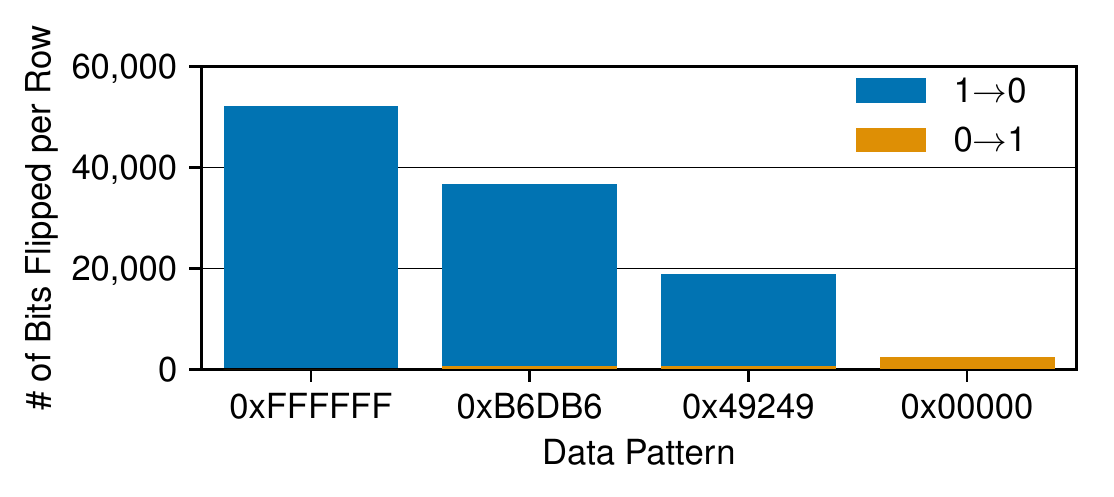}
  \end{center}
  \vspace*{-.5cm}
  \caption{\ffsp Number of bit flips in a representative row for four different data
    patterns: all 1s (0xFFF...), two-third 1s (0xB6DB6D...), one-third 1s (0x492492...),
    all 0s (0x000...).}
  \vspace*{-.5cm}
  \label{fig::full_row_dp}
\end{figure}

To examine this effect, we seeded memory with four data patterns: all
1s, two-thirds 1s (0xB6DB6D...), one-third 1s (0x492492...), and all 0s.
Figure~\ref{fig::full_row_dp} shows the number of bit flips in a victim row
for different data patterns.  This number is directly proportional to the
number of bits seeded with a value of 1: 79.7\% for all 1s, 57\% for
two-thirds 1s, 29.9\% for one-third 1s, and 3.8\% for all 0s.  While bits
can flip in both directions, \emph{most bit flips were seeded with a value
  of 1.}

\begin{figure*}[t]
  \vspace{-.5cm}
  \begin{center}
    \begin{subfigure}[b]{0.5\textwidth}
      \begin{center}
        \includegraphics[height=35mm]{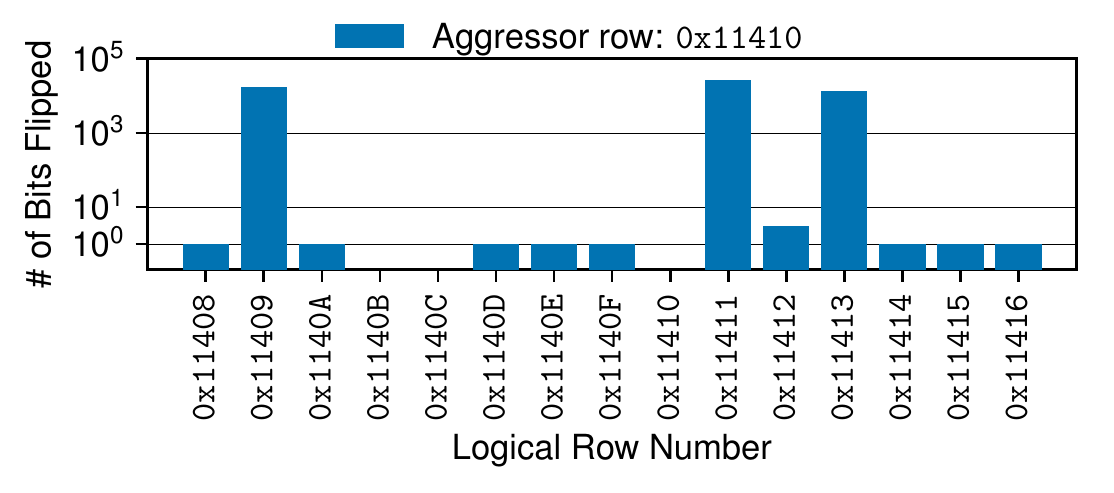}
        \vspace{-.35cm}
        \caption{Victim rows: 0x11411, 0x11409 and 0x11413.}
      \end{center}
    \end{subfigure}%
    ~
    \begin{subfigure}[b]{0.5\textwidth}
      \begin{center}
        \includegraphics[height=35mm]{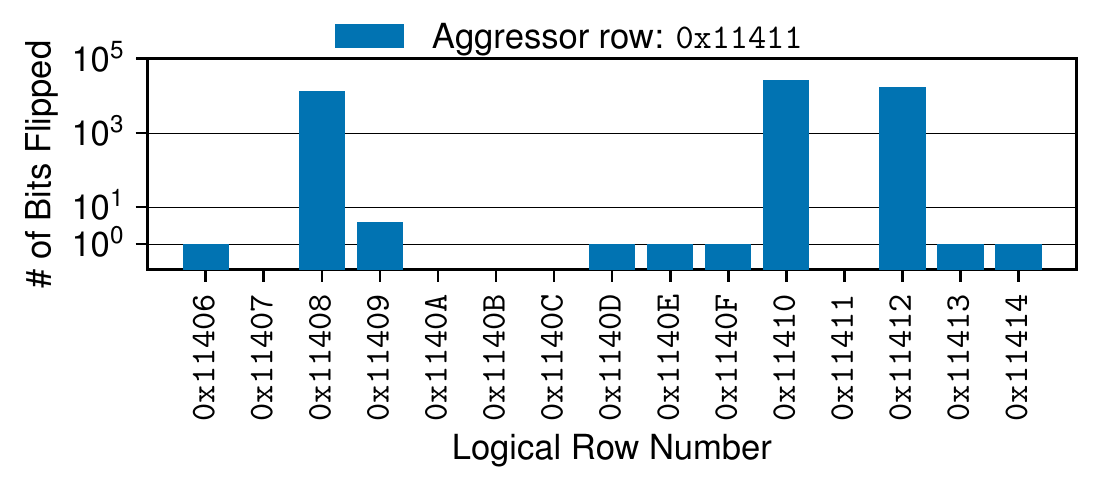}
        \vspace{-.35cm}
        \caption{Victim rows: 0x11410, 0x11408, and 0x11412.}
      \end{center}
    \end{subfigure}
  \end{center}
  \vspace*{-.35cm}
  \caption{\ffsp No. of bit flips on rows with neighboring logical row
    numbers for vendor \#2.}
\vspace*{-.5cm}
\label{fig::re_samsung_rows_0x11410_0x11411}
\end{figure*}

\begin{figure*}[t]
  \begin{center}
    \begin{subfigure}[b]{0.5\textwidth}
      \begin{center}
        \includegraphics[height=35mm]{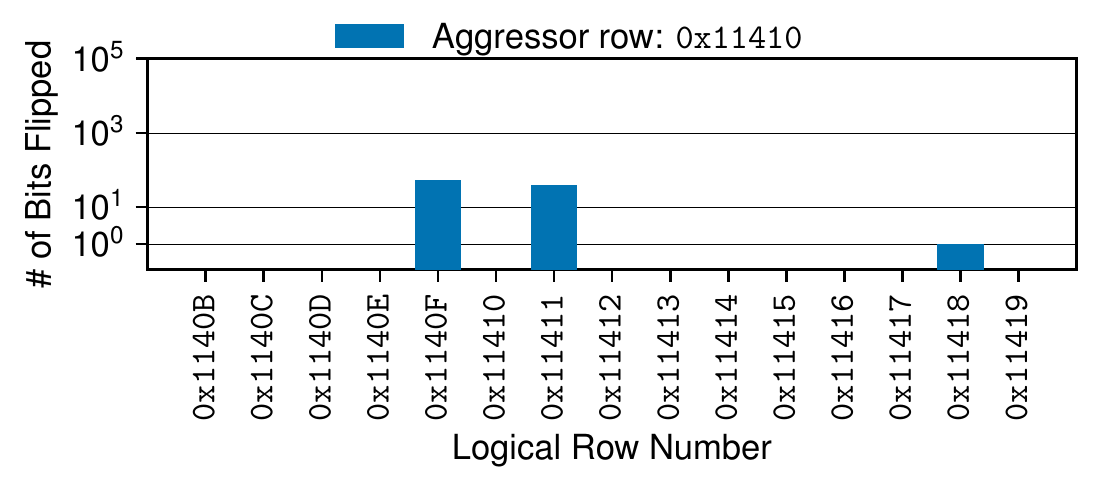}
        \vspace{-.35cm}
        \caption{Victim rows: 0x1140F and 0x11411.}
      \end{center}
    \end{subfigure}%
    ~
    \begin{subfigure}[b]{0.5\textwidth}
      \begin{center}
        \includegraphics[height=35mm]{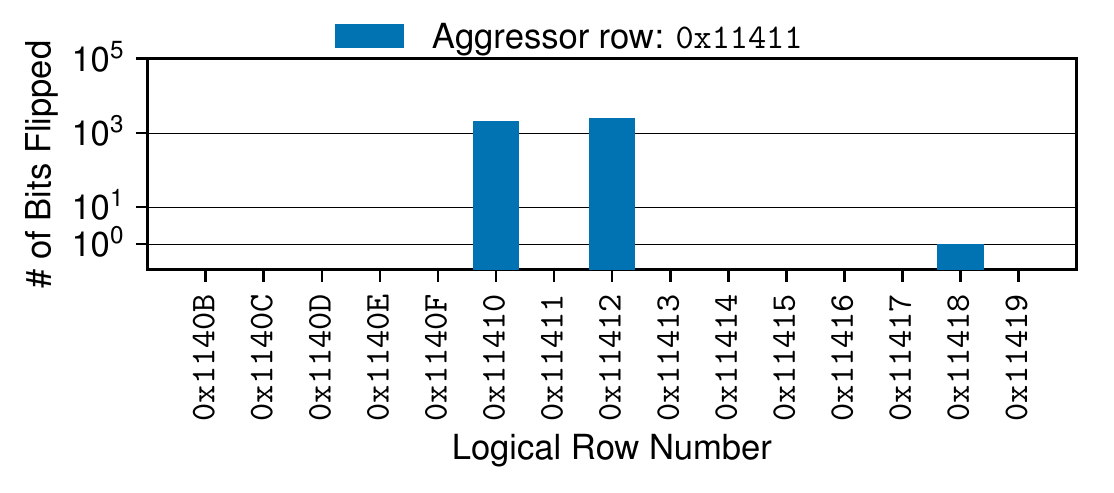}
        \vspace{-.35cm}
        \caption{Victim rows: 0x11410 and 0x11412.}
      \end{center}
    \end{subfigure}
  \end{center}
  \vspace*{-.35cm}
  \caption{\ffsp No. of bit flips on rows with neighboring logical row
    numbers for vendor \#3.}
  \label{fig::re_hynix_rows_0x11410_0x11411}
  \vspace*{-.5cm}
\end{figure*}

\vspace{.2cm}
\noindent
\textbf{DIMMs from vendors \#2 and \#3 have fewer bit flips.}  We repeated
the experiments with DIMMs sourced from the other two hardware vendors. Vendor
\#2 has fewer bit flips per row than vendor \#1.
Figure~\ref{fig::re_samsung_rows_0x11410_0x11411} shows the number of bit
flips for vendor \#2 for the same aggressor rows as before: 0x11410 and
0x11411. When the aggressor row is 0x11410, the whole-row victim (0x11411)
has only 42.1\% of its bits flipped, while the half-row victims (0x11409 and
0x11413) have 25.9\% and 21\% of their bits flipped, respectively.

Further, \emph{the map differs from one vendor to another}.  Vendor \#1's
row 0x11411 has two rows wholly adjacent, rows 0x11410 and 0x11412
(Figure~\ref{fig::re_micron_rows_0x11411_0x11410}).  Instead, for vendor
\#2, this row has whole-row adjacency with row 0x11410 but \emph{only
  half-row} with row 0x11412 and the other half with 0x11408.

Figure~\ref{fig::re_hynix_rows_0x11410_0x11411} shows that vendor \#3 has
far fewer bit flips than both other vendors. When the aggressor is row
\# 0x11410, the adjacent rows have only about 0.08\% of their bits flipped
(roughly two orders of magnitude fewer flips than vendor \#1). When the
aggressor is row \# 0x11411, the adjacent rows have about 3\% of their
bits flipped (one order of magnitude fewer flips than vendor \#1).

To understand whether the lower rate of bit flips for vendors \#2 and \#3
can be attributed to different encoding schemes, we repeated these
experiments with a data pattern of all 0s. We found that even fewer bits
flip.  For all three vendors, most bits in victim rows flip when seeded with a value of 1.

% Micron: 2Rx4 PC4-2666V-RB2-12 MTA36ASF4G72PZ-2G6E1RK  32GB 2Rx4 8Gb (1x)
% Micron: 2Rx4 PC4-2400T-RBB-10 MTA36ASF4G72PZ-2G3B1IK  32GB 2Rx4 8Gb (25nm)
% Samsung: 2Rx4 PC4-2400T-RA1-11-DC0 M393A4K40BB1-CRC0Q S  32GB 2Rx4 8Gb B-die (20nm)
% Hynix: 2Rx4 PC4-2400T-RB1-11 HMA84GR7MFR4N-UH TD AA 1652   32GB 2Rx4 8Gb M-die (25nm)

In order to pinpoint the reasons why vendor \#2 and \#3's DIMMs were more
resilient than vendor \#1's DIMMs, we decoded the part numbers of each
DIMM from each vendor, checking for differences in lithography. We found
that each vendor used a different lithography process; vendor \#1's process
had the highest density, vendor \#2 had lower density, and vendor \#3 had
the lowest one.  These results suggest that increasing DRAM density
increases the probability of flipping bits in a Rowhammer attack. Previous
work also found different rates of bit flips in different
DIMMs~\cite{kim2014disturbance,tatar2018defeating}.

\section{Operational Aspects of DRAM Testing Methodology}
\label{sec::implementation}

% One rank has eight banks.
% One bank has 2GB.
% One row has 8KB
% One bank has 262,144 rows.

Our DRAM testing methodology uses two \emph{clflushopt} instructions in a
loop to \emph{hammer} DRAM.  For each row to test, we use address maps to
identify adjacent rows. We seed the aggressor row with 0s and the victims
with 1s.  If the tested row has two adjacent full-rows, we find two virtual
addresses that map to each of these rows, to seed our two cache line flush
instructions. For each row, we run the test for 128ms, corresponding to
\emph{twice} the duration of a refresh interval.  In this way, we ensure
that our tests span at least one entire refresh interval. At the end of the
test, we check whether the tested row has any of its bits flipped to 0. When
any adjacent row is a half-row, we perform the experiment multiple times,
once for each half-row.

Testing a single bank of our server-class DIMMs using this methodology
takes 11 hours and 36 minutes. Because our DIMMs have 16 banks, testing an
entire DIMM would take about a week. We adopted several
optimizations to better scale our methodology. First, we test several banks
in parallel; we concurrently test eight different banks \emph{on the same
  DIMM} with little interference. Second, rather than testing a row
followed by checking it for bit flips, we batch multiple tests back-to-back
and follow them with a single check at the end.  This reduces the time
needed for the checking step.  With these optimizations in place, our
methodology can test an entire DIMM in less than one day.  In the future,
we plan to further scale up our methodology to simultaneously test multiple
DIMMs by ensuring the tested DIMMs share no channels.

In small-scale experiments with our six server-class DIMMs, we found it
very difficult to flip bits using our testing methodology.  At normal DRAM
refresh rates (with ECC disabled), we observed only two bit
flips on a single DIMM. Despite many additional tests with the same
aggressor rows, we were unable to reproduce these bit flips.  However, at
lower refresh rates, the same DIMM showed hundreds of bit flips (when we
increased the refresh interval by a factor of 3.5x). We were unable
to produce bit flips in our other DIMMs even with a reduced refresh rate.

Our methodology has several limitations we plan to address in future work.

\textbf{Handling TRR.} Some DDR4 DRAM claims it supports Targeted Row
Refresh (TRR)~\cite{ddr4jedec}, a Rowhammer defense in which the DIMM aggressively refreshes
rows under attack. However, researchers have mounted successful Rowhammer
attacks to such memory~\cite{lanteigne2016thirdio, pessl2016drama,
  aga2017good, aichinger2015ddr, cojocar2019ecc}; in these cases, it is
unclear whether TRR is ineffective or not yet enabled.  Because the details
of TRR implementations remain unknown, we designed our methodology to use
the instruction sequence with the highest rate of row activations. If TRR
implementations are reverse engineered, our methodology could be adapted to
use an instruction sequence that \emph{bypasses} TRR defenses while
maximizing the row activation rate.

\textbf{Scaling limitations.}  Unfortunately, our methodology for reverse
engineering row adjacency in a DIMM requires the placement of a hardware
fault injector between a DIMM and its slot. This manual step creates too
much overhead and disruption to be performed at large scale.  Instead, in
practice, we make a simplifying assumption: similar DIMMs sourced from the
same vendor have the same row adjacency map.

Another limitation stems from the choice of using the A14 bit in our
current fault injector design (described in Section~\ref{sec::fi}). For ACT
commands, this bit encodes a row address. Currently, our reverse
engineering methodology cannot use a row whose address has a high bit in
A14. Our methodology therefore tests only half the rows in a bank (those
with a value of '0' for A14). We are currently investigating a more
sophisticated fault injector design that can also work with a high bit in
A14.

\textbf{On generalizing our methodology.} All our experiments were
performed on Intel-based architectures of a cloud provider's compute nodes.
However, cloud providers can use servers based on other types of
architectures, such as AMD or ARM.  Furthermore, in the cloud, DRAM can be
found in many places other than compute nodes, such as storage nodes,
network cards, switches, middleboxes, and so on. Although our results do
not directly transfer to other types of architectures or cloud equipment,
we believe that our methodology can be used directly or adapted to create
worst-case testing conditions.

\textbf{Additional variables.} Our methodology hammers one row for 128ms, a
period of time equal to two refresh intervals. This is the minimal time
interval to ensure that testing one row spans at least one entire refresh
interval from start to end. A more thorough methodology would
determine the duration needed to test a row to declare it \emph{safe}.

In all our experiments, the data values stored in the aggressor rows are
the complement of the values stored in the rest of the bank, a strategy
inspired by previous work~\cite{cojocar2019ecc}. We have not experimented
with storing different data values in the aggressor row.

\section{Related Work}
\label{sec::related}

Many prior works build upon the Rowhammer phenomenon~\cite{onur-date17,
  kim2014disturbance, mutlu2019rowhammer} for both
attacks~\cite{google2015projectzero, google-rh-blackhat,
  kaveh2016flip-feng-shui, veen2016drammer, gruss2016rowhammer-js,
  frigo2018glitch-vu, gruss2018anotherflip, tatar2018throwhammer,
  lipp2018nethammer, xiao2016cloudflops, bosman2016dedup-est-machina,
  qiao2016new, bhattacharya2016curious, jang2017sgx,
  poddebniak2018attacking, aga2017good, tatar2018defeating, pessl2016drama,
  carre2018openssl, cojocar2019ecc, barenghi2018software,
  zhang2018triggering, bhattacharya2018advanced, fournaris2017exploiting,
  ji2019pinpointing, kwong2020rambleed} and defenses~\cite{rh-apple, rh-hp,
  rh-lenovo, rh-cisco, anvil, moin-rowhammer, seyedzadeh2017counter,
  brasser2017can, irazoqui2016mascat, son2017making, gomez2016dram,
  van2018guardion, lee2018twice, bu2018srasa, bains2015row, bains14d,
  bains14c, bains16refresh-cmd, greenfield15condition-monitoring,
  konoth2018zebram, izzo2017reliably, gong2018memory, jones2017holistic,
  kline2017sustainable, schilling2018pointing, vig2018rapid}. Few of these
works provide insight into the difficulty behind mounting an attack on a
real system, and none develop a methodology for thoroughly profiling a DIMM
for Rowhammer vulnerability. We discuss the most closely related works to
ours. A detailed survey of a very large number of Rowhammer related works
can be found in~\cite{mutlu2019rowhammer}.

\textbf{Rowhammer Testing Platforms.}  Kim et al.~\cite{kim2014disturbance}
first studied the Rowhammer disturbance effect on DDR3 using a custom
FPGA-based memory controller~\cite{softmc, softmc-safarigithub} that
directly interfaces with and sends DDR3 commands to DRAM devices.
They~\cite{kim2014disturbance} also study Rowhammer on Intel and AMD platforms.
\cite{francis2018raspberry} implements an OS for testing memory devices on
Raspberry Pi platforms.  Drammer~\cite{veen2016drammer, drammer-github,
  drammer-app-github}, an open-source Android app, tests mobile
devices for Rowhammer and gathers data from users to characterize how
widespread Rowhammer is.  MemTest86~\cite{rh-passmark}, software that tests
DRAM for many types of reliability issues, added Rowhammer testing.
\cite{yim2016methodology} presents a methodology for injecting a Rowhammer
attack that is largely complementary to ours because it focuses on ways to
place a victim page into a vulnerable memory location.
\cite{yim2016methodology} also emulates Rowhammer failures and evaluates
them by injecting errors into an OS kernel. While these previous works
provide insight into studying Rowhammer in real DRAM devices, they either do
not create worst-case testing conditions or do not work in end-to-end
systems like our cloud servers.

\textbf{Physical Row Adjacency.}  Other works attempt to reverse engineer
the DRAM address mapping with various techniques, such as
side-channels~\cite{pessl2016drama, jung2016crosshair}, software fault
injection~\cite{tatar2018defeating}, or hardware fault
injection~\cite{cojocar2019ecc, intel2019mei}.
Section~\ref{sec::methodology::challenges::adjacency} discusses their
shortcomings in depth.

A number of other works~\cite{yang2019trap, yun2018study, park2016root,
  ryu2017overcoming} sidestep the address translation issue and study the
Rowhammer phenomenon directly at the circuit level with simulations.
Unfortunately, these works do not allow us to study Rowhammer
characteristics on real devices.

\textbf{Optimal Rowhammer Access Pattern.} Previous works used a plethora
of instruction sequences to test for
Rowhammer~\cite{kim2014disturbance,google2015projectzero,
  xiao2016cloudflops, bosman2016dedup-est-machina,qiao2016new,
  bhattacharya2016curious, gruss2016rowhammer-js,pessl2016drama,
  kaveh2016flip-feng-shui, veen2016drammer,lanteigne2016thirdio,
  jang2017sgx, aga2017good, gruss2017guide,gruss2018anotherflip,
  tatar2018defeating}. These works measure an instruction sequence's
efficiency by quantifying the number of Rowhammer failures it induces on a set
of DIMMs.  In contrast, our work (1) characterizes an instruction
sequence's rate of ACTs, (2) describes and evaluates the factors that
prevent the sequence from achieving the optimal ACT rate, and (3) proposes a
new instruction sequence that is near-optimal on Skylake, a platform
commonly used in modern cloud servers.

\section{Conclusions}
\label{sec::conclusions}

This paper presents an end-to-end methodology for rigorously evaluating the
susceptibility of cloud servers to Rowhammer attacks. Our methodology
creates worst-case DRAM testing conditions by overcoming two main
challenges: (1) generating the highest rate of row activation commands to
DRAM, and (2) testing rows whose cells are physically adjacent inside a
DRAM device.  Cloud providers can adapt our techniques to test their
infrastructure in practice and determine whether a Rowhammer attack can
cause DRAM bits to flip.

Applying our methodology to multiple generations of servers from a major
cloud provider produced noteworthy results. First, none of the instruction
sequences used in previous work to mount Rowhammer attacks create
near-optimal row activation rates. Second, we constructed a new instruction
sequence that generates a near-optimal rate of row activations on Skylake
and Cascade Lake. This instruction sequence leverages microarchitectural
side-effects to create row activations and uses no explicit memory
accesses. Third, we designed and implemented a fault injector capable of
reverse engineering row adjacency inside any DRAM device on a DIMM.
Finally, we used our fault injector to reverse engineer the physical
adjacency of DIMMs sourced from three major DRAM vendors.  Our results show
that logical rows do not always map linearly inside DRAM devices.

%XXX: Camera-ready
% To do list for camera-ready
% - replace ``row adjacency in a DRAM device'' with ``row adjacency at the
% DIMM-level''

\vspace{.2cm} \textbf{Acknowledgments.}  We would like to thank our
shepherd, Yinqian Zhang, our point-of-contact reviewer, Christian Rossow,
and the anonymous reviewers for their feedback on our paper. We received
huge support and feedback from the Azure Cloud Server Infrastructure and
Azure Networking teams, especially from Kushagra Vaid, Dave Maltz, Tanj
Bennett, Rama Bhimanadhuni, Tim Cowles, Kelley Dobelstein, Sriram Govindan,
Terry Grunzke, Sushant Gupta, Bryan Kelly, Phyllis Ng, Jitu Padhye, Joseph
Piasecki, Andrew Putnam, Shachar Raindel, Mark Shaw, and Lit Wong.  We
thank Barbara Aichinger from FuturePlus for her valuable help with the
FS2800 bus analyzer.  We are grateful to Kaveh Razavi for answering many
questions on Rowhammer and sending us DIMMs from his lab. We thank Stephen
Dahl for his help setting up our hardware lab.  Finally, we greatly
appreciate the overall support of Victor Bahl.

% Balance paper's last page of text
\balance

%\bibliographystyle{acm}
%\bibliography{rh}
{
  % \lsstyle % ENABLE ME IF YOU WISH TO SEE YOUR FIRSTBORN KILLED BEFORE YOUR VERY EYES

  \footnotesize
  \let\OLDthebibliography\thebibliography
  \renewcommand\thebibliography[1]{
    \OLDthebibliography{#1}
    \setlength{\parskip}{0pt}
    \setlength{\itemsep}{0pt}
  }
  \bibliographystyle{IEEEtranS}
  \bibliography{rh}
}

%Note: ZombieLoad seems to imply that the line-buffer might cache
%non-temporal loads.

%\input{appendix}
\end{document}